\documentclass[prd, 11pt,nofootinbib, superscriptaddress, preprintnumbers,floatfix]{revtex4}

\usepackage[utf8]{inputenc}
\usepackage{amsmath,amssymb}
\usepackage{graphicx}
\usepackage{subfigure}
\usepackage{hyperref}
\usepackage{nicefrac}
\usepackage{xr}
\usepackage{footnote}
\usepackage{booktabs}
\usepackage{import}
\usepackage{braket}
\usepackage{multirow}
\usepackage{here}
\usepackage{color}
\usepackage{dsfont}
\usepackage{placeins}
\usepackage[normalem]{ulem}
\newcommand{\stkout}[1]{\ifmmode\text{\sout{\ensuremath{#1}}}\else\sout{#1}\fi}

\newcommand{\eq}{\begin{eqnarray}}
\newcommand{\en}{\end{eqnarray}}

\begin{document}

\title{Energy shift of the three-particle system in a finite volume}

\newcommand\bn{HISKP and BCTP, Rheinische Friedrich-Wilhelms Universit\"at Bonn, 53115 Bonn, Germany}
\newcommand\ucas{School of Physical Sciences, University of Chinese
Academy of Sciences, Beijing 100049, China}
\newcommand\da{Institut f\"ur Kernphysik, Technische Universit\"at Darmstadt,
64289 Darmstadt, Germany\\ and ExtreMe Matter Institute EMMI, GSI Helmholtzzentrum f\"ur Schwerionenforschung,\\ 64291 Darmstadt, Germany}
\newcommand\ju{Institute for Advanced Simulation (IAS-4),
Institut f\"ur Kernphysik  (IKP-3) and\\ J\"ulich Center for Hadron Physics,
Forschungszentrum J\"ulich,\\ D-52425 J\"ulich, Germany}
\newcommand\tb{Ivane Javakhishvili Tbilisi State University, 0186 Tbilisi, Georgia}

\preprint{MITP/18-092}

\author{Jin-Yi~Pang}\affiliation{\bn}
\author{Jia-Jun Wu}\affiliation{\ucas}\affiliation{\bn}
\author{Hans-Werner Hammer}\affiliation{\da}
\author{Ulf-G.~Mei{\ss}ner}\affiliation{\bn}\affiliation{\ju}\affiliation{\tb}
\author{Akaki~Rusetsky}\affiliation{\bn}

\begin{abstract}

Using the three-particle quantization condition recently obtained in the 
particle-dimer framework, the finite-volume energy shift of the
two lowest three-particle scattering states is derived up to and including
order $L^{-6}$.
Furthermore, assuming that a stable dimer
exists in the infinite volume, the shift for the lowest particle-dimer 
scattering state is obtained up to and including order $L^{-3}$. 
The result for the lowest three-particle state agrees with the results
from the literature, and the result for the lowest particle-dimer state
reproduces the one obtained by using the L\"uscher equation.

\end{abstract}

\date{\today}

\maketitle

\section{Introduction}

Recent years have seen a rapid progress in developing the formalism for the 
analysis of the lattice data in the three-particle sector~\cite{Polejaeva:2012ut,Meissner:2014dea,Guo:2016fgl,Guo:2017ism,Briceno:2012rv,Hansen:2014eka,Hansen:2015zga,Hansen:2015zta,Hansen:2016fzj,Hansen:2016ync,Briceno:2017tce,Kreuzer:2010ti,Kreuzer:2009jp,Kreuzer:2008bi,Kreuzer:2012sr,Sharpe:2017jej,pang1,pang2,Meng:2017jgx,Doring:2018xxx,Romero-Lopez:2018rcb,Mai:2017bge,Briceno:2018mlh,Briceno:2018aml,Mai:2018djl,Guo:2018xbv,Guo:2018ibd,Blanton:2019igq} (for the recent review, see Ref.~\cite{Hansen:2019nir}). One of the main goals of
this development was the derivation of the so-called quantization condition, 
i.e., the equation that relates the finite-volume energy spectrum with the infinite-volume observables~\cite{Polejaeva:2012ut,Briceno:2012rv,Hansen:2014eka,pang2,Mai:2017bge}. In particular, in our previous papers~\cite{pang1,pang2,Doring:2018xxx} we have proposed an approach to the problem that is based on the use of
 the non-relativistic effective theory and the particle-dimer picture in a 
finite volume (a very similar approach was proposed in Refs.~\cite{Kreuzer:2008bi,Mai:2017bge}).
Within this approach, a tower of particle-dimer couplings with 
increasing inverse mass dimension, describing the three-particle force, is fit to the 
lattice spectrum in the three-particle sector. The physical observables are
then constructed through solving the scattering equations in the infinite 
volume, using the input determined on the lattice. It has been shown~\cite{pang2}
 that this approach is conceptually equivalent to other approaches in the
 literature.

In case of two particles, the infinite-volume spectrum consists of a 
continuum and (possibly) bound states below the two-particle threshold. 
In a finite volume, this all translates into a discrete spectrum. If there 
were no interactions, the momenta of the particles in a box of size $L$ would 
be quantized, according to 
${\bf p}_i=2\pi{\bf n}_i/L,~{\bf n}_i\in \mathbb{Z}^3$ for $i=1,2$, and the spectrum
would be given by a sum of the free energies of two particles. In the presence
of interactions, the energy levels are displaced from their free values.
In this case, the spectrum consists of the levels that in the infinite-volume
  limit end up in the
  continuum (we refer to these levels as to the scattering states), and the levels that
are continuously transformed into the bound states in this limit. 
This displacement of the {\em scattering states} can be calculated in a systematic
expansion in $1/L$, by 
using the two-particle quantization condition
(L\"uscher equation)~\cite{Luscher:1986pf}. The final result
for the lowest scattering state takes 
the form~\cite{Huang,Wu,Tan,Beane}
\eq\label{eq:E2}
\Delta E_2=E_2(L)-2m&=&\frac{4\pi a}{mL^3}\,\biggl\{1+c_1\biggl(\frac{a}{\pi L}\biggr)
+c_2\biggl(\frac{a}{\pi L}\biggr)^2
+c_3\biggl(\frac{a}{\pi L}\biggr)^3
+\frac{2\pi ra^2}{L^3}-\frac{\pi a}{m^2L^3}\biggr\}
\nonumber\\[2mm]
&+&O(L^{-7})\, ,
\en
where, for simplicity, we display the formula for the displacement
of the ground-state energy levels of two identical particles with the
mass $m$. In this formula, $a$ and $r$ denote the two-body S-wave scattering length 
and the effective range, respectively, and $c_1,c_2,c_3$ are known numerical 
constants. The last term is the relativistic correction, which vanishes, 
as $m\to\infty$. The mixing to other partial waves does not enter at this order in $1/L$.

The spectrum of the three-particle system is richer. If the
two-body interaction is attractive and sufficiently strong, bound state(s)
(or stable dimer(s))
may be formed and, apart of the three-particle scattering states,
we shall have
particle-dimer scattering states as well\footnote{Following the
    terminology widely used in the literature, throughout this paper we use the
    name ``dimer'' in two different contexts. Namely, in this paragraph, a ``stable dimer'' is merely equivalent to the notion of the two-body bound state (a stable particle). In a more general context, a ``dimer'' is an auxiliary field with given quantum numbers that enables one to transform the three-particle equations to a simpler form.
    The existence of a stable particle is no more required (of course, if such a particle
    exists, the two interpretations are equivalent). 
    We expect that the terminology used here does not lead to a confusion.    
    } In particular, as shown in Ref.~\cite{Doring:2018xxx}, the finite-volume energy levels in such a system display characteristic avoided level crossings at those values of $L$, 
where the free energies of the three-particle and particle-dimer systems coincide.

As in the two-particle case, the energy levels of an interacting three-body
system in a finite volume are shifted from their free values. The expression
for the energy shift of the lowest
three-particle scattering state is known in the 
literature  up to and including $O(L^{-6})$ (see, e.g., Refs.~\cite{Huang,Wu,Tan,Beane,Hansen:2015zta,Hansen:2016fzj}) and even up to and including $O(L^{-7})$,
see Ref.~\cite{Detmold:2008gh}. The formula up to and including $O(L^{-6})$
takes the form 
\eq
    \Delta E_3 =E_3(L)-3m&=&
  \frac{12 \pi a}{m L^3} \biggl\{ 1 + d_1 \biggl(\frac{a}{\pi L}\biggr)
    + d_2 \left(\frac{a}{\pi L} \right)^2
    + \frac{3\pi a}{ m^2L^3}  + \frac{6\pi r a^2}{L^3}
\nonumber\\[2mm]
    &+& \left(\frac{a}{\pi L} \right)^3 d_3 \ln \frac{m L}{2\pi}   \biggr\}- \frac{D}{48 m^3 L^6}  + O(L^{-7})\,, \label{eq:E3}
\en
where $d_1,d_2,d_3$ are the known numerical constants, the quantity $D$
contains the contribution of the three-body force and the fourth term inside the brackets, 
containing inverse power of $m$, is a relativistic correction. The expression for the 
excited three-particle levels, as well as the particle-dimer levels
(in case these exist), to the best of our knowledge, are not available
in the literature. Our present work fills this gap.
In particular, we derive the finite volume shift of the lowest
three-particle and particle-dimer scattering states 
in a systematic expansion in powers of $1/L$ (including logarithms),
using the approach of Ref.~\cite{pang2}. Note that we do not consider
three-particle bound states (trimers) which were discussed
in Refs.~\cite{Meissner:2014dea,Hansen:2016ync,pang1} and thus
in the following refer to the lowest scattering state as ``ground state''
for convenience.

The justification of our enterprise is twofold. First and foremost, we provide
  closed analytic expressions for the energy shift of different levels
  in a finite volume, which can be
  easily fitted to the lattice data. Moreover, these expressions are written
  directly in terms of infinite-volume observables (like two-body effective-range expansion
  parameters of the three-body threshold amplitude). On the contrary, fitting the numerical solution of the quantization condition directly to the lattice data is much more complicated
  and less transparent. The second argument in favor of our study is that deriving the
  perturbative shifts of different levels is a powerful check of the approach, since, e.g.,
  the result for the ground state has been already available in the literature.

In short, our result for the three-particle scattering states has the following form:
\eq
E_3(L)-E_3^{\sf free}(L)=\frac{P_0}{L^3}+\frac{P_1}{L^4}+\frac{P_2}{L^5}
+\frac{P_3}{L^6}\,\ln\frac{mL}{2\pi}+\frac{P_4}{L^6}+O(L^{-7})\, ,
\en
where $E_3^{\sf free}(L)$ is the energy of three free particles in a box (the ground state or the first excited state in the center-of-mass frame), whereas the coefficients
$P_0,\ldots,P_4$ are different for different levels and different irreps of the octahedral group. As mentioned before, these coefficients depend only on the infinite-volume physical observables. Moreover, we have found the formula of the ground-state shift of a particle
and a tightly bound stable dimer:
\eq
E_{1d}(L)-E_{1d}^{\sf free}(L)=\frac{{\cal A}_d}{L^3}+O(L^{-4})\, ,
\en
where ${\cal A}_d$ denotes the particle-dimer scattering amplitude at threshold.

An explicit result for the
energy shift of the three-body ground state is given in
Eq.~(\ref{eq:final-ground}), while the results for
first excited state are given in Eq.~(\ref{eq:final-excited})
for the $A_1^+$ representation and Eq.~(\ref{eq:final-excited-E})
for the $E^+$ representation. Compact, easy to use formulas
with numerical coefficients for these
energy shifts are given in Appendix~\ref{app:num-res}.
Finally, the shift for the lowest
particle-dimer level is given in Eq.~(\ref{eq:particle-dimer-shift}).
In the course of our work, we also obtain a simple and transparent closed 
expression for the three-particle contribution of the three-body scattering amplitude at threshold.
The lattice practitioners, who are only interested in the final results, are directly
referred to the Appendix~\ref{app:num-res}.

The plan of our paper is rather straightforward. First, in Section~\ref{sec:identity} we write down a basis matrix identity that enables one to derive the perturbative expansion of the energy shift in powers of $1/L$. Next, in Section~\ref{sec:groundstate} we derive the formula for the shift of the three-particle 
ground state up to and including $L^{-6}$. In Section~\ref{sec:dimer} we address the calculation of the particle-dimer levels to leading order.
 In Section~\ref{sec:excited}, we 
 calculate the first excited three-particle energy level shift up to and including $L^{-6}$.
 Section~\ref{sec:comparison} is dedicated to the comparison to
 previous results in the literature. 
 Finally, Section~\ref{sec:concl} contains our conclusions and an
 outlook. All technical details are relegated to the appendices.
  
\section{Transforming the quantization condition}
\label{sec:identity}

The quantization condition for three identical spinless bosons, given in
Refs.~\cite{pang2,Doring:2018xxx}, is obtained under certain assumptions, namely:
(i) the particles obey the non-relativistic dispersion law and (ii) only S-wave interactions
are taken into account. The work to go beyond these assumptions is already in progress,
but for the time being we restrict ourselves to the existing framework.

Under these assumptions, the quantization condition takes the
form $\det{\cal A}=0$, where the matrix ${\cal A}$ is defined in momentum space:
\eq\label{eq:Apq}
{\cal A}_{pq}=8\pi\tau_L^{-1}({\bf p};E)L^3\delta_{{\bf p}{\bf q}}-8\pi Z({\bf p},{\bf q};E)
\doteq (\tau^{-1}-Z)_{pq}\, .
\en
Here, the three-momenta ${\bf p},{\bf q}$ take the discrete values,
${\bf p}=2\pi{\bf n}/L,~{\bf n}=\mathbb{Z}^3$ (similarly for ${\bf q}$).
A cutoff $\Lambda$
is imposed, demanding that the matrix is finite, with ${\bf p},{\bf q}<\Lambda$.
Note also that writing down the quantization condition in the matrix notation,
we do not show the factors $L^3$ explicitly. An overall factor $8\pi$ is included
in the definition of $\tau^{-1}_{pq}$ and $Z_{pq}$.
The sum over each momentum index $p$ (not boldface) always implies the factor $1/L^3$,
whereas $\delta_{pq}=L^3\delta_{{\bf p}{\bf q}}$.
Further,
\eq\label{eq:tauL}
8\pi\tau_L^{-1}({\bf p};E)&=&p^*\cot\delta(p^*)+S({\bf p},(p^*)^2)\, ,
\nonumber\\[2mm]
S({\bf p},(p^*)^2)&=&-\frac{4\pi}{L^3}\sum_{\bf l}\frac{1}{{\bf p}^2+{\bf p}{\bf l}+{\bf l}^2-mE}\, ,
\en
and $p^*$ is the magnitude of the 
relative momentum of the pair in the rest frame, with
\eq
(p^*)^2=mE-\frac{3}{4}\,{\bf p}^2\, .
\en
In Eq.~(\ref{eq:tauL}) the momentum sum is implicitly 
regularized by using dimensional regularization, leading eventually to the pertinent
L\"uscher zeta-function\footnote{Namely, in order to regularize this expression, we subtract and add the same expression, where the sum is replaced by an integral. The integral is then calculated within the dimensional regularization.}. Further, $\delta(p^*)$ is the S-wave
phase shift in the two-particle subsystem.
The effective range expansion reads
\eq
p^*\cot\delta(p^*)=-\frac{1}{a}+\frac{1}{2}\,r(p^*)^2+O((p^*)^4)\, ,
\label{eq:ERE}
\en
where $a,r$ are the two-body scattering length and the effective
range, respectively.

The quantity $Z$ in Eq.~(\ref{eq:Apq}) denotes the kernel of the Faddeev
equation. It contains the one-particle exchange diagram, as well as the local
 term,
corresponding to the particle-dimer interaction (three-particle force).
We shall see that, up to and including terms of order $L^{-6}$,
it suffices to retain the non-derivative
coupling only, which is described by a single constant $H_0(\Lambda)$.
The kernel then takes the form
\eq\label{eq:Z}
Z({\bf p},{\bf q};E)=\frac{1}{-mE+{\bf p}^2+{\bf q}^2+{\bf p}{\bf q}}
+\frac{H_0(\Lambda)}{\Lambda^2}\, .
\en
The dependence of $H_0(\Lambda)$ on the cutoff is such that the infinite-volume
scattering amplitude is cutoff-independent. In a finite volume, this ensures
the cutoff-independence of the spectrum.

After giving a short summary of the approach, introduced in
Refs.~\cite{pang1,pang2,Doring:2018xxx}, we next turn to the derivation of the
perturbative expansion of the energy level shifts on the basis of the quantization
condition. Here, we first note that the equation $\det{\cal A}=0$, which defines all
energies, is very difficult to handle directly. For this reason, we shall
first derive the basic matrix identity, which will then allow us to carry
out the expansion of the energy shift in powers of $1/L$. This identity gives the
determinant as a product of two factors: the determinant of a matrix
with certain row(s)/column(s) erased, and the remainder. The crucial point is that,
in the case of free particles,
the different rows/columns of the matrix ${\cal A}$ correspond to the
well-defined free levels. Then,
the properly ``chopped'' determinant in the interacting case
should not have a zero, corresponding to a given shifted free level (here, one assumes
that the shift is perturbative in $1/L$ and that it is much smaller than the distance between
levels). Consequently, all information about a given level is contained in the remainder,
which has a much simpler form and where the unnecessary information about all other
levels is already factored out.

In order to demonstrate  how this factorization works in practice, consider a
generic matrix ${\cal A}$ and select an arbitrary value of the index, $r$. Let
${\cal A}_{/pq}$ denote a matrix, which is obtained from ${\cal A}$ by crossing out the
row $p$ and the column $q$. Expanding
the determinant of the matrix ${\cal A}$ in co-factors, we get
\eq
\det{\cal A}={\cal A}_{rr}\det{\cal A}_{/rr}-\sum_{k,l\neq r}(-1)^{k+l}
{\cal A}_{rk}{\cal A}_{lr}\det{\cal A}_{/rk/lr}\, .
\en
Taking into the fact that ${\cal A}_{/rk/lr}={\cal A}_{/rr/lk}$ and that the
inverse of a matrix $A_{/rr}$ is equal to its conjugate divided by its determinant:
\eq
({\cal A}_{/rr}^{-1})_{kl}=(-1)^{k+l}\det{\cal A}_{/rr/lk}(\det{\cal A}_{/rr})^{-1}\, ,
\en
we get:
\eq
\det{\cal A}=\biggl({\cal A}_{rr}-\sum_{k,l\neq r}{\cal A}_{rk}({\cal A}_{/rr})^{-1}_{kl}
{\cal A}_{lr}\biggr)\det{\cal A}_{/rr}\, .
\en
Hence, from the quantization condition $\det{\cal A}=0$ it follows that
the quantity in brackets, which does not contain determinants anymore, should vanish\footnote{Of course, this condition gives only part of the roots of the whole
  determinant -- those, which are {\em not} given by the solution of the equation
  $\det{\cal A}_{/rr} = 0$.
  The root we are interested in -- the one that lies in the vicinity of a given unperturbed
  energy level -- is  among these roots.}.

In our case, ${\cal A}_{pq} = (\tau^{-1}-Z)_{pq}=(\tau^{-1}[1-\tau Z])_{pq}$. Taking into
account the fact that $\tau$ is a diagonal matrix, i.e., $\tau_{pq}=\delta_{pq}\tau_p$, we get:
\eq\label{eq:series}
{\cal A}_{rr}
=(\tau^{-1}-Z)_{rr}=\sum_{k,l\neq r}{\cal A}_{rk}({\cal A}_{/rr})^{-1}_{kl}
{\cal A}_{lr}
&=&
\sum_{k,l\neq r}Z_{rk}(1-\tau Z)^{-1}_{kl}\tau_lZ_{lr}
\nonumber\\[2mm]
&=&\sum_{k\neq r}Z_{rk}\tau_kZ_{kr}
+\sum_{k,l\neq r}Z_{rk}\tau_kZ_{kl}\tau_lZ_{lr}+\cdots
\en
This matrix identity can be used to produce a systematic expansion of the
energy shift in powers of $1/L$, as we shall see in the following. Here, we just note that
the r.h.s. of Eq.~(\ref{eq:series}) is nothing but the multiple-scattering expansion
of the particle-dimer scattering amplitude in a finite volume,
starting from the second term. At the
end, everything will sum up in the infinite-volume exact particle-dimer amplitude plus
a well-defined set of the finite-volume
corrections that emerge from the first few terms in this expansion.

\section{Three particle ground state energy shift}
\label{sec:groundstate}

\subsection{Solving the quantization condition}

In order to calculate the ground state energy shift, we have to single out the
matrix element ${\cal A}_{rr}$, where $r$ corresponds to the state with ${\bf p}=0$.
Further, the quantity $E$ in the quantization condition corresponds to the energy
above the three free particle threshold, i.e., to the energy shift $\Delta E_3(L)$ in this
particular case.

It can be seen from Eq.~(\ref{eq:series}) that $E$ is of order $L^{-3}$.
We arrive at this conclusion neglecting the r.h.s. of this equation altogether. It can then be
checked {\it a posteriori} that taking the r.h.s. into account gives only the $O(L^{-1})$ corrections
to the leading-order result. Further, since
the momenta are of order $L^{-1}$, the Taylor expansion of the energy denominators in powers of $q_0^2$
is justified and, for the quantity $\tau_L^{-1}({\bf 0};E)$, one obtains
\eq\label{eq:tau0}
8\pi\tau_L^{-1}({\bf 0};E)&=&-\frac{1}{a}+\frac{1}{2}\,rq_0^2+\cdots
-\frac{4\pi}{L^3}\sum_{\bf l}\frac{1}{{\bf l}^2-q_0^2}
\nonumber\\[2mm]
&=&-\frac{1}{a}+\frac{2\pi^2}{L^2}\,r\kappa^2
+\frac{1}{\pi L}\,\frac{1}{\kappa^2}-\frac{1}{\pi L}\,{\cal I}-\frac{1}{\pi L}\,
\kappa^2{\cal J}-\frac{1}{\pi L}\,\kappa^4{\cal K}+\cdots\, ,
\en
where $q_0^2=mE$ and $\kappa=Lq_0/(2\pi)$.
 The constants ${\cal I},~{\cal J},~{\cal K}$ are given in Appendix~\ref{app:sums}, see also Ref.~\cite{Beane}.
Recalling that at leading order $q_0^2=O(L^{-3})$ and hence $\kappa^2=O(L^{-1})$, it is easy to verify that the neglected
terms in Eq.~(\ref{eq:tau0}) are of order of $L^{-4}$ and higher.

The expansion of the quantity $\tau_L^{-1}({\bf p};E)$ at a  nonzero value of the
momentum ${\bf p}$ can be obtained similarly (one needs only leading terms in this
expansion):
\eq
8\pi\tau_L^{-1}({\bf p};E)=-\frac{1}{a}
+\sqrt{\frac{3}{4}\,{\bf p}^2}
-\frac{1}{\pi L}\,\tilde{\cal I}({\bf n})
+\cdots\, ,
\en
where ${\bf p}=2\pi {\bf n}/L$ and
\eq
\tilde{\cal I}({\bf n})=
\lim_{N\to\infty}\biggr\{
\sum_{\bf j}^N\frac{1}
{{\bf j}^2+{\bf n}^2+{\bf n}{\bf j}}-4\pi N
+2\pi^2\sqrt{\frac{3}{4}\,{\bf n}^2}\biggr\}\, .
\en
Note that the above equation, unlike Eq.~(\ref{eq:tau0}),
 does not contain singular terms, proportional 
to $\kappa^{-2}$. 

Using Eq.~(\ref{eq:series})
the three-body quantization condition is written as:
\eq\label{eq:quant}
8\pi\tau_L^{-1}\biggl({\bf 0};\frac{q_0^2}{m}\biggr)-
\frac{8\pi}{L^3}\,Z\biggl({\bf 0},{\bf 0};\frac{q_0^2}{m}\biggr)
-\frac{8\pi}{L^3}\,\Delta=0\, ,
\en
where,
\eq
\Delta=\sum\limits_{k\neq r}Z_{rk}\tau_kZ_{kr}
+\sum\limits_{k,l\neq r}Z_{rk}\tau_kZ_{kl}\tau_lZ_{lr}+\cdots\, .
\en
The quantity $\Delta$ does not contain singular terms either.
Expanding in powers of $\kappa^2$, we get
\eq\label{eq:Delta}
L^{-3}\Delta=\frac{\Delta^{(0)}}{L^2}+\frac{\Delta^{(1)}}{L^3}\,\ln\frac{mL}{2\pi}
+\frac{\Delta^{(2)}}{L^3}+\frac{\Delta^{(3)}}{L^2}\,\kappa^2+O(L^{-4})\, .
\en
Taking into account Eqs.~(\ref{eq:Z}), (\ref{eq:tau0}) and (\ref{eq:Delta}), 
the three-body quantization condition, Eq.~(\ref{eq:quant}),
can be solved with respect to $\kappa^2$ by iterations. The result,
up to and including terms of order $L^{-6}$,
is given by:
\eq\label{eq:Eshift}
\kappa^2&=&\frac{3a}{\pi L}\biggl\{1-\biggl(\frac{a}{\pi L}\biggr){\cal I}
+\biggl(\frac{a}{\pi L}\biggr)^2\biggl({\cal I}^2-3{\cal J}-\frac{(2\pi)^3}{a}\,\Delta^{(0)}\biggr)-\frac{8\pi a}{L^3}\,\Delta^{(1)}\ln\frac{mL}{2\pi}
+\frac{6\pi a^2r}{L^3}
+\frac{Y}{L^3}\biggr\}\, ,
\nonumber\\
\en
where
\eq\label{eq:X}
Y=\biggl(\frac{a}{\pi}\biggr)^3\biggl(-{\cal I}^3+9{\cal I}{\cal J}-9{\cal K}
+\frac{(2\pi)^3}{a}\,2{\cal I}\Delta^{(0)}\biggr)
-8\pi a\biggl(\frac{H_0(\Lambda)}{\Lambda^2}+\Delta^{(2)}+\frac{3a}{\pi}\,
\Delta^{(3)}\biggr)\, .
  \en
  Hence, in order to complete the calculation, it remains to evaluate the quantities
  $\Delta^{(0)},\ldots,\Delta^{(3)}$. To this end, let us consider the terms, entering in $\Delta$,
  separately.

  \subsection{One iteration}
\label{sec:one-iteration}

The crucial point in the derivation of the energy shift is that the 
leading-order corrections (in $1/L$) emerge from the first two terms in the
expression of $\Delta$. In all other terms, up to and including order $L^{-6}$,
one could simply perform the infinite-volume limit, replacing sums by integrals.
In the following, we shall illustrate this statement by explicit calculations,
writing $\Delta=S_1+S_2+\ldots$ and starting
 from the first term in the expansion:
  \eq\label{eq:S1}
  S_1=\sum_{k\neq 0}Z_{0k}\tau_kZ_{k0}
  =\frac{8\pi}{L^3}\sum_{{\bf k}\neq{\bf 0}}^\Lambda
  \biggl(\frac{1}{{\bf k}^2-mE}+\frac{H_0(\Lambda)}{\Lambda^2}\biggr)^2
  \frac{-a}{1-a\sqrt{\dfrac{3}{4}\,{\bf k}^2}+\dfrac{a}{\pi L}\,\tilde{\cal I}({\bf n})}\, ,
  \en
  where ${\bf k}=2\pi{\bf n}/L$. 

The leading corrections arise from the most infrared-singular terms.
In order to single out these corrections, we first note that it is safe to expand the above
expression in powers of $q_0^2=mE$ and $\dfrac{a}{\pi L}\,\tilde{\cal I}({\bf n})$.
namely, the quantity $\tilde{\cal I}({\bf n})$ is uniformly bound from above for
all values of ${\bf n}$ (in difference to $\sqrt{{\bf k}^2}$, which also appears in the denominator).
The result of the expansion is given by
  \eq
S_1&=&\frac{8\pi}{L^3}\sum_{{\bf k}\neq{\bf 0}}^\Lambda
  \biggl(\frac{1}{{\bf k}^2}+\frac{H_0(\Lambda)}{\Lambda^2}\biggr)^2
  \frac{-a}{1-a\sqrt{\dfrac{3}{4}\,{\bf k}^2}}
  \nonumber\\[2mm]
&+&  \frac{8\pi}{L^3}\sum_{{\bf k}\neq{\bf 0}}^\Lambda
\frac{2q_0^2}{{\bf k}^4}\,  \biggl(\frac{1}{{\bf k}^2}+\frac{H_0(\Lambda)}{\Lambda^2}\biggr)
  \frac{-a}{1-a\sqrt{\dfrac{3}{4}\,{\bf k}^2}}
\nonumber\\[2mm]
&-& \frac{8\pi}{L^3}\sum_{{\bf k}\neq{\bf 0}}^\Lambda
\frac{a\tilde{\cal I}({\bf n})}{\pi L}\,
\biggl(\frac{1}{{\bf k}^2}+\frac{H_0(\Lambda)}{\Lambda^2}\biggr)^2
  \frac{-a}{1-a\sqrt{\dfrac{3}{4}\,{\bf k}^2}}+\cdots
  \nonumber\\[2mm]
  &=&S_1'+S_1''+S_1'''+\cdots\, .
  \en
  Below, we shall consider all these terms separately and
  start from the first one, writing $S_1'=(S_1'-\bar S_1')+\bar S_1'$, where $\bar S_1'$ is the most singular
  piece of $S_1'$, as ${\bf k}\to 0$:
  \eq\label{eq:S11}
  \bar S_1'=\frac{8\pi}{L^3}\sum_{{\bf k}\neq{\bf 0}}^\Lambda
  \frac{1}{{\bf k}^4}\,(-a)\biggl\{1+a\sqrt{\dfrac{3}{4}\,{\bf k}^2}\biggr\}\, .
  \en
  It is immediately seen that the difference $(S_1'-\bar S_1')$ goes as ${\bf k}^{-2}$,
  when ${\bf k}\to 0$. 
  Further, for the evaluation of the sums in Eq.~(\ref{eq:S11}) we shall use the formulae given in
  Ref.~\cite{Hansen:2015zta}:
  \eq
  \frac{1}{L^3}\sum_{\bf p}\frac{f({\bf p}^2)}{{\bf p}^2}
  &=&\int\frac{d^3{\bf p}}{(2\pi)^3}\,\frac{f({\bf p}^2)}{{\bf p}^2}
  +\frac{{\cal I}f(0)}{(2\pi)^2L}-\frac{f'(0)}{L^3}+\cdots\, ,
  \nonumber\\[2mm]
  \frac{1}{L^3}\sum_{\bf p}\frac{g({\bf p}^2)}{{\bf p}^4}
  &=&\frac{L{\cal J}g(0)}{(2\pi)^4}
  + \int\frac{d^3{\bf p}}{(2\pi)^3}\,\frac{g({\bf p}^2)-g(0)}{{\bf p}^4}+\cdots\, ,
\nonumber\\[2mm]
\frac{1}{L^3}\sum_{\bf p}\frac{h({\bf p}^2)}{{\bf p}^6}
&=&\frac{L^3{\cal K}h(0)}{(2\pi)^4}+\cdots\, .
\en
The most infrared-singular piece in Eq.~(\ref{eq:S11}) is the one containing
$1/{\bf k}^4$. To calculate this contribution, we shall demonstrate a trick that will be
systematically used below:
\eq\label{eq:1eps}
\frac{1}{L^3}\,\sum_{{\bf k}\neq{\bf 0}}^\Lambda\frac{1}{{\bf k}^4}
&=&\frac{L{\cal J}}{(2\pi)^4}-\lim_{\varepsilon\to 0}\int_\Lambda^\infty
\frac{d^3{\bf k}}{(2\pi)^3}\,
\frac{1}{({\bf k}^2+\varepsilon^2)^2}
\nonumber\\[2mm]
&=&\lim_{\varepsilon\to 0}\biggl[\int^\Lambda\frac{d^3{\bf k}}{(2\pi)^3}\,
\frac{1}{({\bf k}^2+\varepsilon^2)^2}-\frac{1}{8\pi\varepsilon}\biggr]
+\frac{L{\cal J}}{(2\pi)^4}\, .
\en  
The physical meaning of introducing the quantity $\varepsilon$ is to shift the total
energy of the particle-dimer system $E=-\varepsilon^2/m$ slightly {\em below}
threshold. The particle-dimer amplitude at zero initial and final momenta is singular
at threshold corresponding to $\varepsilon=0$. Subtracting the quantity
$(8\pi\varepsilon)^{-1}$ from the amplitude eliminates the singularity, and the limit $\varepsilon\to 0$ can be performed in the remainder. We shall do the
same procedure in all terms in the perturbative series. The key property of these series is
that the number of the diagrams, leading to the singular behavior at threshold, is finite,
even if the perturbative series for the amplitude contain infinitely many terms. Consequently,
one may find all singular terms in a closed form. In a finite volume, exactly these singular
diagrams lead to the leading contributions to the energy shift.

In order to demonstrate the above statement, consider, for example, a less singular
term in $S_{1}^{'}$, containing one power of the coupling $H_0(\Lambda)$:
\eq
\frac{1}{L^3}\,\sum_{{\bf k}\neq{\bf 0}}^\Lambda \frac{2}{{\bf k}^2}\,
\frac{H_0(\Lambda)}{\Lambda^2}=
\lim_{\varepsilon\to 0}\int^\Lambda\frac{d^3{\bf k}}{(2\pi)^3}\,
\frac{2}{{\bf k}^2+\varepsilon^2}\,
\frac{H_0(\Lambda)}{\Lambda^2}+\cdots\, .
\en
This expression is not singular at $\varepsilon\to 0$ and behaves as $L^0$ at large $L$,
whereas the integral in Eq.~(\ref{eq:1eps}) has a singular piece of order
$\varepsilon^{-1}$ that translates into an $O(L)$ term at large $L$.

Continuing in the same manner with the expansion in  Eq.~(\ref{eq:1eps}) we calculate the
next most singular contribution to Eq.~(\ref{eq:S11}): 
\eq
\frac{1}{L^3}\,\sum_{{\bf k}\neq{\bf 0}}^\Lambda\frac{1}{|{\bf k}|^3}
&=&\frac{1}{(2\pi)^3}\,\biggl(\sum_{{\bf j}\neq{\bf 0}}^{\Lambda L/2\pi}
\frac{1}{|{\bf j}|^3}
-\int_1^{\Lambda L/2\pi}\frac{d^3{\bf j}}{|{\bf j}|^3}
+\int_1^{\Lambda L/2\pi}\frac{d^3{\bf j}}{|{\bf j}|^3}\biggr)
\nonumber\\[2mm]
&=&\frac{1}{(2\pi)^3}\,\biggl(4\pi\ln\frac{\Lambda L}{2\pi}
+{\cal L}+O(L^{-1})\biggr)\, ,
\en
where the quantity ${\cal L}$ is given in Appendix~\ref{app:sums}, see also
Appendix~\ref{app:integrals}.

Using the same trick, this expression can be rewritten as
\eq
\frac{1}{L^3}\,\sum_{{\bf k}\neq{\bf 0}}^\Lambda\frac{1}{|{\bf k}|^3}
&=&\frac{2}{\sqrt{3}}\,\lim_{\varepsilon\to 0}\biggl[\int^\Lambda\frac{d^3{\bf k}}{(2\pi)^3}\,
\frac{1}{({\bf k}^2+\varepsilon^2)^2}\,
\sqrt{\frac{3}{4}\,{\bf k}^2+\varepsilon^2}+\frac{\sqrt{3}}{4\pi^2}\,\ln\frac{\varepsilon}{m}
  \biggr]
\nonumber\\[2mm]
&+&\frac{1}{2\pi^2}\,\ln\frac{mL}{2\pi}+\frac{1}{2\pi^2}\,
\biggl(\frac{1}{2}\,(1-\ln 3)+\frac{\sqrt{3}\pi}{18}\biggr)
  +\frac{1}{(2\pi)^3}\,{\cal L}+\cdots\, .
\nonumber\\ 
 \en
  There are no more singular terms in $S_1'$, given by Eq.~(\ref{eq:S11}) and, hence,
  the summation can be directly replaced by the integration in the remaining terms.
  Further, the quantities $S_1''$ and $S_1'''$ can be evaluated directly:
  \eq
  S_1''&=&-\frac{aq_0^2L^3}{4\pi^5}\,{\cal K}+\cdots\, ,
  \nonumber\\[2mm]
  S_1'''&=&\frac{a^2}{2\pi^4}\,{\cal G}+\cdots\, ,
  \en
  where the ellipses stand for the sub-leading terms in $L$ and the quantity ${\cal G}$
  in given in Appendix~\ref{app:sums}, see also Appendix~\ref{app:integrals}.

  Putting all pieces together, we finally obtain
  \eq
  S_1&=&\lim_{\varepsilon\to 0}\biggr[
  \int^\Lambda\frac{d^3{\bf k}}{(2\pi)^3}\biggl(\frac{1}{{\bf k}^2+\varepsilon^2}+\frac{H_0(\Lambda)}{\Lambda^2}\biggr)^2
  \frac{-8\pi a}{1-a\sqrt{\dfrac{3}{4}\,{\bf k}^2+\varepsilon^2}}+\frac{a}{\varepsilon}
  -\frac{2\sqrt{3}a^2}{\pi}\,\ln\frac{\varepsilon}{m}\biggr]
\nonumber\\[2mm]
  &-&\frac{aL{\cal J}}{2\pi^3}
  -\frac{2\sqrt{3}a^2}{\pi}\,\ln\frac{mL}{2\pi}
  -\frac{2\sqrt{3}a^2}{\pi}\,
\biggl(\frac{1}{2}\,(1-\ln 3)+\frac{\sqrt{3}\pi}{18}\biggr)
  -\frac{a^2\sqrt{3}}{2\pi^2}\,{\cal L}
\nonumber\\[2mm]
  &+&\frac{a^2}{2\pi^4}\,{\cal G}
  -\frac{aq_0^2L^3}{4\pi^5}\,{\cal K}
+\cdots\, .
  \en
  Note that the divergence-subtracted threshold particle-dimer amplitude appears in the first line of this equation.

  \subsection{Two iterations}
\label{sec:two-iterations}

The method, used above, can be applied to the second term in $\Delta$, which is written down as:
  \eq\label{eq:S2}
  S_2&=&\frac{(8\pi)^2}{L^6}\,\sum_{{\bf k}\neq{\bf 0}}^\Lambda\sum_{{\bf q}\neq{\bf 0}}^\Lambda
  \biggl(\frac{1}{{\bf k}^2}+\frac{H_0(\Lambda)}{\Lambda^2}\biggr)
  \frac{-a}{1-a\sqrt{\dfrac{3}{4}\,{\bf k}^2}}\,
  \biggl(\frac{1}{{\bf k}^2+{\bf q}^2+{\bf k}{\bf q}}
  +\frac{H_0(\Lambda)}{\Lambda^2}\biggr)
\nonumber\\[2mm]
  &\times&\frac{-a}{1-a\sqrt{\dfrac{3}{4}\,{\bf q}^2}}\,
  \biggl(\frac{1}{{\bf q}^2}+\frac{H_0(\Lambda)}{\Lambda^2}\biggr)\, .
  \en
  Note that here we consistently drop the terms proportional to $q_0^2$ and $\tilde{\cal I}({\bf n})$, since these do not contribute at the accuracy we are working.
  There is only one singular diagram at this order $S_2=(S_2-\bar S_2)+\bar S_2$, where
     \eq\label{eq:S21}
  \bar S_2=a^2\frac{(8\pi)^2}{L^6}\,\sum_{{\bf k}\neq{\bf 0}}^\Lambda\sum_{{\bf q}\neq{\bf 0}}^\Lambda
  \frac{1}{{\bf k}^2}\,
  \frac{1}{{\bf k}^2+{\bf q}^2+{\bf k}{\bf q}}\,
  \frac{1}{{\bf q}^2}
  =\frac{a^2}{\pi^4}\,\biggl\{
  \frac{8\pi^4}{3}\,\ln\frac{\Lambda L}{2\pi}+2{\cal Q}
\biggr\}\, .
    \en
where the quantity ${\cal Q}$ is given in Appendix~\ref{app:sums}
(see Appendix~\ref{app:integrals}
  for the derivation of this formula).

    On the other hand, we can use our trick again and calculate the integral
    \eq\label{eq:tI0}
    I_\varepsilon=a^2\int^\Lambda \frac{d^3{\bf k}}{(2\pi)^3}\,\int^\Lambda \frac{d^3{\bf q}}{(2\pi)^3}\,\frac{1}{{\bf k}^2+\varepsilon^2}\,
  \frac{1}{{\bf k}^2+{\bf q}^2+{\bf k}{\bf q}+\varepsilon^2}\,
  \frac{1}{{\bf q}^2+\varepsilon^2}=\frac{a^2}{24\pi^2}\,\ln\frac{\Lambda}{\varepsilon}+\frac{a^2}{(2\pi)^6}\,{\cal I}_0\, ,
  \en
  where ${\cal I}_0\simeq -375.754658$ denotes a numerical constant. The derivation of the above formula,
  as well as an explicit expression for this constant is contained in Appendix~\ref{app:integrals}. Since all other terms in the expansion of $S_2$, given by the Eq.~(\ref{eq:S2}), are non-singular, one immediately gets
  \eq
  S_2&=&\lim_{\varepsilon\to 0}\biggl[\int^\Lambda \frac{d^3{\bf k}}{(2\pi)^3}\,\int^\Lambda \frac{d^3{\bf q}}{(2\pi)^3}\,
  \biggl(\frac{1}{{\bf k}^2+\varepsilon^2}+\frac{H_0(\Lambda)}{\Lambda^2}\biggr)
  \frac{-8\pi a}{1-a\sqrt{\dfrac{3}{4}\,{\bf k}^2+\varepsilon^2}}
\nonumber\\[2mm]
  &\times&  \biggl(\frac{1}{{\bf k}^2+{\bf q}^2+{\bf k}{\bf q}+\varepsilon^2}
  +\frac{H_0(\Lambda)}{\Lambda^2}\biggr)
\frac{-8\pi a}{1-a\sqrt{\dfrac{3}{4}\,{\bf q}^2+\varepsilon^2}}\,
  \biggl(\frac{1}{{\bf q}^2+\varepsilon^2}+\frac{H_0(\Lambda)}{\Lambda^2}\biggr)
  +\frac{8a^2}{3}\,\ln\frac{\varepsilon}{m}\biggr]
\nonumber\\[2mm]
  &+&\frac{8a^2}{3}\,\ln\frac{mL}{2\pi}+\frac{a^2}{\pi^4}\,
  (2{\cal Q}-{\cal I}_0)\, .
  \en
  Note that there are no singular terms at higher orders in the iteration series, so one can stop here and write down an expression for the quantity $\Delta$ defined in Eq.~(\ref{eq:Delta}).

\subsection{An expression for the quantity $\Delta$}
  
In order to write down a closed expression for the quantity $\Delta$, it is convenient to
define the divergence-subtracted threshold particle-dimer amplitude in the following
manner:
\begin{enumerate}
\item
  Consider the particle-dimer scattering amplitude ${\cal M}\biggl({\bf 0},{\bf 0};-\frac{\varepsilon^2}{m}\biggr)$, which is a solution of the Faddeev equation in the infinite volume with the kernel given by $Z$.
\item
  Taking into account that $Z\biggl({\bf 0},{\bf 0};-\frac{\varepsilon^2}{m}\biggr)
  =\dfrac{1}{\varepsilon^2}+\dfrac{H_0(\Lambda)}{\Lambda^2}$,
  see Eq.~(\ref{eq:Z}), one can define the divergence-subtracted amplitude
  $\hat{\cal M}$ as
  \eq
  {\cal M}\biggl({\bf 0},{\bf 0};-\frac{\varepsilon^2}{m}\biggr)
  =\frac{{\cal M}_{-2}}{\varepsilon^2}+\frac{{\cal M}_{-1}}{\varepsilon}
  +{\cal M}_0\ln\frac{\varepsilon}{m}+\hat{\cal M}+\cdots\, ,
  \en
  where the coefficients ${\cal M}_{-2},{\cal M}_{-1},{\cal M}_0$ are adjusted so that
  $\hat{\cal M}$ does not depend on $\varepsilon$ in the limit $\varepsilon\to 0$, and
  the ellipses stand for the terms which vanish in this limit.
\end{enumerate}

Now, we can easily read off the expressions for $\Delta^{(0)},\ldots,\Delta^{(3)}$ from the above formulae:
\eq
\Delta^{(0)}&=&-\frac{a{\cal J}}{2\pi^3}\, ,
\nonumber\\[2mm]
\Delta^{(1)}&=&-\frac{2\sqrt{3}a^2}{\pi}+\frac{8a^2}{3}\, ,
\nonumber\\[2mm]
\frac{H_0(\Lambda)}{\Lambda^2}+\Delta^{(2)}&=&\hat{\cal M}+\frac{a^2}{(2\pi)^3}\,{\cal B}\, ,
\nonumber\\[2mm]
\Delta^{(3)}&=&-\frac{a}{\pi^3}\,{\cal K}\, ,
\en
where ${\cal B}$ is a combination of many numerical factors, emerging in different equations:
\eq\label{eq:B}
{\cal B}=
  -16\pi^2\sqrt{3}
\biggl(\frac{1}{2}\,(1-\ln 3)+\frac{\sqrt{3}\pi}{18}\biggr)
  -4\pi\sqrt{3}{\cal L}
+\frac{4}{\pi}\,{\cal G}
+\frac{8}{\pi}\,
  (2{\cal Q}-{\cal I}_0)\, .
  \en

\subsection{Final expression for the ground-state energy shift at $O(L^{-6})$}

After substituting into Eqs.~(\ref{eq:Eshift}) and (\ref{eq:X}), we finally get
\eq\label{eq:final-ground}
\kappa^2&=&\frac{3 a}{\pi L}\biggl\{1-\biggl(\frac{a}{\pi L}\biggr){\cal I}
+\biggl(\frac{a}{\pi L}\biggr)^2\biggl({\cal I}^2+{\cal J}\biggr)
+\biggl(\frac{a}{\pi L}\biggr)^316\pi^3\biggl(\sqrt{3}-\frac{4\pi}{3}\biggr)
\ln\frac{mL}{2\pi}
\nonumber\\[2mm]
&+&\frac{6\pi a^2r}{L^3}
+\frac{Y}{L^3}\biggr\}\, ,
\en
where
\eq
Y=\biggl(\frac{a}{\pi}\biggr)^3\biggl(-{\cal I}^3+{\cal I}{\cal J}+15{\cal K}-\pi {\cal B}\biggr)
-8\pi a\hat{\cal M}\, .
\en
In order to simplify the comparison of our result with the ones from the literature, it is convenient to introduce the notation
${\cal R}=({\cal G}-\pi^2\sqrt{3}{\cal L})/2\simeq-32.60560475$. Our result can be then rewritten as
\eq
Y=\biggl(\frac{a}{\pi}\biggr)^3\biggl(-{\cal I}^3+{\cal I}{\cal J}+15{\cal K}-
8(2{\cal Q}+{\cal R})\biggr)
-8\pi a\biggl(\hat{\cal M}+\frac{a^2}{8\pi^3}\,\delta\biggr)\, ,
  \en
where the quantity
\eq\label{eq:delta}
\delta=  -16\pi^2\sqrt{3}
\biggl(\frac{1}{2}\,(1-\ln 3)+\frac{\sqrt{3}\pi}{18}\biggr)
-\frac{8}{\pi}\,{\cal I}_0 \simeq 887.65392
  \en
contains only infinite-volume integrals.

\section{Shift of the lowest particle-dimer scattering level}
\label{sec:dimer}

If the two-body scattering length is positive and large, there exists a shallow two-body
bound state (a dimer). The set of the asymptotic states in the three-body problem
in the infinite volume then includes the states, describing a dimer and a spectator,
separated by a large distance and do not interacting with each other. The energy of such
an asymptotic state is $E=E_d+\dfrac{{\bf p}^2}{2(2m+E_d)}+\dfrac{{\bf p}^2}{2m}$,
where $2m+E_d$ is the mass of a dimer is the rest frame and ${\bf p}$ is
the momentum of the spectator in the rest frame of three particles.
In case of a large scattering length, we have
$E_d=-1/(ma^2)+\ldots$, where the ellipses stand for the terms emerging from higher
orders in the effective range expansion.
One expects that
these infinite-volume scattering levels obtain shifts in a finite volume, just similar to what the three-particle
levels do. It will be seen below that our quantization condition allows one to calculate
these shifts as well.

In this paper, we restrict ourselves to the calculation of the shift of the 
particle-dimer ground state at order $L^{-3}$. This corresponds to the spectator momentum
${\bf p}=0$. The energy of the state is given by $E=E_d+\Delta_d$, where $\Delta_d$ denotes
the finite-volume shift. The quantization condition is given by, cf. Eq.~(\ref{eq:quant}):
\eq\label{eq:quant-dimer}
8\pi\tau_L^{-1}({\bf 0};E)-\frac{8\pi}{L^3}\,Z({\bf 0},{\bf 0};E)-\frac{8\pi}{L^3}\,\Delta({\bf 0},{\bf 0};E)=0\, ,
\en
where the quantity $\Delta$ contains the terms with multiple iterations, see the
previous section.
Further, since $E_d$ is below the two-particle threshold,
\eq
8\pi\tau_L^{-1}({\bf 0};E)=8\pi K^{-1}({\bf 0};E)+\sqrt{-mE}+\cdots\, ,
\en
where the ellipses stand for the exponentially suppressed terms and
\eq
8\pi K^{-1}({\bf 0};E) =-\frac{1}{a}+\frac{1}{2}\,r(mE)+\cdots
\en
denotes the infinite-volume two-body $K$-matrix in the rest frame.

Solving the quantization condition for $\Delta_d$ iteratively, one can replace $E$ by $E_d$
in $Z(E)+\Delta(E)$, arriving at the {\em threshold amplitude for the particle-dimer
  scattering}
\eq
{\cal M}_d=Z({\bf 0},{\bf 0};E_d)+\Delta({\bf 0},{\bf 0};E_d)\, .
\en
Note that, in contrast to the threshold amplitude, introduced in the previous section,
this quantity does not contain infrared singularities, since $E_d\neq 0$ from the beginning.
The solution of the quantization condition to the leading order is given by
\eq\label{eq:Luescher-dimer}
\Delta_{d}=\frac{{\cal Z}}{L^3}\,{\cal M}_d\, ,
\en
where
\eq
{\cal Z}&=&\frac{-8\pi}{\dfrac{m}{2\sqrt{-mE_d}}-(8\pi K^{-1})'}\, ,
\nonumber\\[2mm]
(8\pi K^{-1})'&=&\frac{d}{dE}\,\biggl(8\pi K^{-1}({\bf 0};E)\biggr)\biggr|_{E=E_d}\, .
\en
It is straightforwardly seen that the quantity  $\mathcal{Z}$ coincides
with the wave function renormalization constant for the dimer field. Indeed,
in the normalization, used in this paper, the two-point function of the dimer
field is given by\footnote{The normalization of the dimer propagator can be 
easily read off from Eqs.~(329) and (333) of Ref.~\cite{Braaten:2004rn}, where
the effective-range expansion in the two-body interaction is carried out at leading order only. Namely, comparing Eq.~(333) from that paper with our Eq.~(\ref{eq:Z}), one sees that the overall normalization of the particle-dimer amplitudes
differs by a factor $g_2^2/4$, where $g_2$ denotes the leading-order two-body coupling. This fixes the normalization in Eq.~(\ref{eq:dimer-prop}), where
exactly the same multiplicative factor has to be removed, as compared
to Eq.~(329) of Ref.~\cite{Braaten:2004rn}.}
\eq\label{eq:dimer-prop}
D(E,{\bf 0})=\frac{8\pi}{8\pi K^{-1}(E)-\sqrt{-mE-i\varepsilon}}\to
\frac{\mathcal{Z}}{E_d-E}+\cdots\, . 
\en
Consequently, the quantity ${\cal A}_d=\mathcal{Z}{\cal M}_d$ is just the particle-dimer
scattering amplitude at threshold, and Eq.~(\ref{eq:Luescher-dimer})
can be rewritten as
\eq
\Delta_d=\frac{1}{L^3}\,{\cal A}_d\, ,
\label{eq:particle-dimer-shift}
\en
which exactly reproduces the two-particle L\"uscher equation~(\ref{eq:E2}),
where the dimer is considered as an elementary particle. This result, of course, does not come as
a surprise. In fact, if a dimer is deeply bound and thus can be considered as elementary,
the L\"uscher formula will hold in higher orders in $1/L$ as well. The particle-dimer
scattering length, the effective range, etc., which appear in this formula, are then related to
the parameters of the underlying theory in the same way, as in the infinite volume.
However, the case of a loosely bound dimer is less trivial. The leading term in $1/L$ expansion is still given by the L\"uscher formula, but the sub-leading terms
will contain corrections, which can be attributed to its finite spatial extension. The calculation of these corrections is, however, beyond the scope of the present paper.

The main difference of the particle-dimer energy shift $\Delta_d$ with the shift of the
three-particle state is that here the quantity ${\cal M}_d$ appears at leading order already.
Hence, the particle-dimer shift will be more sensitive to the value of the three-body
coupling $H_0(\Lambda)$. This fact can be used for the extraction of the latter quantity
from the lattice data.

\section{The shift of the excited three-particle level}
\label{sec:excited}

\subsection{Singular contributions}

The calculation of the shift of the excited levels proceeds along the similar
path as for the ground state level, but some modifications are needed. In
order to understand the origin of the emerging problems, consider, for 
instance, the shift of the unperturbed level with the particles having momenta
${\bf p}_1=(0,0,1)$, ${\bf p}_2=(0,0,-1)$ and ${\bf p}_3=(0,0,0)$ (all momenta
are given in units of $2\pi/L$). There is no unique assignment for the 
spectator momentum in this case: the magnitude of the spectator momenta can
be equal to either $0$ or $1$. Let us recall now that, in case of the ground 
state, the matrix ${\cal A}$, considered in Section~\ref{sec:identity},
contained singular contributions, proportional to $1/\kappa^2$, only in the diagonal matrix
element ${\cal A}_{rr}$, corresponding to the spectator momentum $|{\bf p}|=0$.
 The quantization condition could be then readily rewritten in form     
of Eq.~(\ref{eq:series}). In case of the excited levels the situation is 
different -- now, the singular terms are contained in a block, which corresponds
to the different assignments of the spectator momenta. In the case, considered
above, this will be a $2\times 2$ block, whose entries correspond to momenta
$|{\bf p}|=0,1$. Hence, the technique, described in Section~\ref{sec:identity},
needs to be amended.

For simplicity, we explicitly describe the calculations for the first excited
three-particle level, where the momenta are chosen as above. The technique
is, however, completely general and can be applied for higher excited states 
as well. Moreover, already the first excited level will be split between
different irreps of the octahedral group, which was not the case for the 
ground-state -- there, the shifted level emerged in the irrep $A_1^+$ only.
Hence, one has to
 project the quantization condition onto the different irreps of the
octahedral groups as described in Ref.~\cite{Doring:2018xxx}. Here, we 
use the notations and terminology from that paper. In particular, we refer to
the set of lattice momenta, which can be obtained from a single reference momentum
${\bf p}_0$ by applying
the elements of the octahedral group
${\bf p}=g{\bf p}_0,~g\in O_h$, as to {\em shells}. In particular,
$|{\bf p}|=0$ and $|{\bf p}|=1$ correspond to the shells $r=1$ and $r=2$, respectively.
The quantization condition, projected onto the irrep $\Gamma$, takes the form
\eq\label{eq:quant-cond}
\det\biggl(8\pi\tau_L^{-1}(r;E)\delta_{rs}\delta_{\sigma\rho}
-\frac{8\pi\theta^{1/2}(r)\theta^{1/2}(s)}{GL^3}\, Z^\Gamma_{\rho\sigma}(r,s;E)\biggr)=0\, ,
\en
where $r,s$ number the shells, $\sigma,\rho$ label the basis vectors in the irrep $\Gamma$, $\theta(s)$ is the multiplicity of a given shell (the number of the independent vectors
in this shell), $G=48$ is the total number of elements in $O_h$ and
\eq
Z^\Gamma_{\rho\sigma}(r,s;E)
=\sum_{g\in O_h}(T^\Gamma_{\sigma\rho}(g))^*Z(g{\bf p}_0(r),{\bf k}_0(s);E)
\en
is the projection of the driving term $Z$ onto the irrep $\Gamma$. 
Here, the set of matrices $T^\Gamma_{\sigma\rho}(g)$ forms an irreducible
matrix representation $\Gamma$ of the octahedral group $O_h$.

Below, we shall first derive the expression for the energy shift for the 
irrep $A_1^+$. This irrep is one-dimensional and therefore one may suppress the indices $\sigma,\rho$. Other irreps will be considered later. The expansion for
the quantity $\tau_L^{-1}$ on the first shell, corresponding to ${\bf p}=0$,
is given by
(cf. with Eq.~(\ref{eq:tau0})):   
\eq\label{eq:tauL1}
8\pi\tau_L^{-1}(1;E)&=&-\frac{1}{a}+\frac{1}{2}\,rq_0^2+\cdots-\frac{4\pi}{L^3}\,
\sum_{\bf l}\frac{1}{{\bf l}^2-q_0^2}
=-\frac{1}{a}+\frac{2\pi^2}{L^2}\,r\kappa^2+\cdots
\nonumber\\[2mm]
&-&\frac{6}{\pi L}\,\frac{1}{1-\kappa^2}
-\frac{1}{\pi L}\,{\cal I}_1+\frac{1}{\pi L}\,{\cal J}_1(1-\kappa^2)
-\frac{1}{\pi L}\,{\cal K}_1(1-\kappa^2)^2+\cdots\, ,
\en
where $\kappa^2=1$ corresponds to the unperturbed first
excited level. Further, the numerical factors ${\cal I}_1,~{\cal J}_1,~{\cal K}_1$
 are given in Appendix~\ref{app:sums}.

For the expansion on the second shell, let us choose the reference momentum
${\bf p}=\dfrac{2\pi}{L}\,{\bf j}_0$ with ${\bf j}_0=(0,0,1)$. We get:
\eq\label{eq:tauL2}
8\pi\tau_L^{-1}(2;E)&=&-\frac{1}{a}+\frac{1}{2}\,r\biggl(q_0^2-\frac{3}{4}\,{\bf p}^2\biggr)+\cdots-\frac{4\pi}{L^3}\,
\sum_{\bf l}\frac{1}{{\bf l}^2+{\bf p}{\bf l}+{\bf p}^2-q_0^2}
\nonumber\\[2mm]
&=&-\frac{1}{a}+\frac{2\pi^2}{L^2}\,r\biggl(\kappa^2-\frac{3}{4}\biggr)
+\cdots
-\frac{2}{\pi L}\,\frac{1}{1-\kappa^2}
\nonumber\\[2mm]
&-&\frac{1}{\pi L}\,{\cal I}_2+\frac{1}{\pi L}\,{\cal J}_2(1-\kappa^2)
-\frac{1}{\pi L}\,{\cal K}_2(1-\kappa^2)^2+\cdots\, ,
\en
where ${\cal I}_2,~{\cal J}_2,~{\cal K}_2$ are defined in Appendix~\ref{app:sums}.

The expansion on the higher shells does not contain singular terms, proportional
to $(1-\kappa^2)^{-1}$. Hence, it suffices to retain fewer terms in this expansion
for $s>2$:
\eq
8\pi\tau_L^{-1}(s;E)=-\frac{1}{a}+\sqrt{\frac{3}{4}\,{\bf p}^2-mE}+
\cdots-\frac{1}{\pi L}\,
{\cal I}({\bf j}_s)
\,,
\en
where
\eq
{\cal I}({\bf j}_s)
=\lim_{N\to\infty}
\biggl(\sum_{\bf n}^N\frac{1}{{\bf n}^2+{\bf n}{{\bf j}_s}
  +{\bf j}^2_s-1}-4\pi N+2\pi^2\sqrt{\frac{3}{4}\,{\bf j}^2_s-1}\biggr)\, ,
\en
and the reference momentum on the shell $s$ is given by
${\bf p}=\dfrac{2\pi}{L}\,{\bf j}_s$.

Next, we consider the projection of the driving term $Z$ onto the irrep $A_1^+$.
This is a trivial representation $T(g)=1,~g\in O_h$. The projection then yields:
\eq\label{eq:ZA1}
\frac{8\pi\theta(1)}{GL^3}\,\sum_{g\in O_h}Z(g{\bf 0},{\bf 0};E)
&=&\frac{2}{\pi L}\,\frac{1}{-\kappa^2}+\frac{8\pi}{L^3}\,
\frac{H_0(\Lambda)}{\Lambda^2}\, ,
\nonumber\\[2mm]\frac{8\pi\sqrt{\theta(1)\theta(2)}}{GL^3}\,\sum_{g\in O_h}Z(g{\bf 0},{\bf q};E)
&=&\frac{2\sqrt{6}}{\pi L}\,\frac{1}{1-\kappa^2}+\frac{8\sqrt{6}\pi}{L^3}\,\frac{H_0(\Lambda)}{\Lambda^2}\, ,
\nonumber\\[2mm]
\frac{8\pi\theta(2)}{GL^3}\,\sum_{g\in O_h}Z(g{\bf p},{\bf q};E)
&=&
\frac{2}{\pi L}\,\frac{1}{1-\kappa^2}
+\frac{8}{\pi L}\,\frac{1}{2-\kappa^2}
+\frac{2}{\pi L}\,\frac{1}{3-\kappa^2}+\frac{48\pi}{L^3}\,\frac{H_0(\Lambda)}{\Lambda^2}\, .
\en
Here, ${\bf p},{\bf q}$ denote two arbitrary vectors on the shell $s=2$. Like the
quantity $\tau_L^{-1}$, the projected quantity $Z$ on higher shells does not develop
singularities, as expected. Thus, an explicit projection of $Z$ will not be needed for these
shells.

\subsection{Solving the quantization condition for the excited states}

In the ground state, only the diagonal matrix element of ${\cal A}$, corresponding to
the shell $s=1$, contained a singularity. One could then directly apply the method
described in Section~\ref{sec:identity}. However, as we have seen, the structure
of the matrix ${\cal A}$ in the vicinity of the free excited state energy is different, namely
\eq
{\cal A}=\left(
  \begin{array}{c c | c c}
    {\cal A}_{11} & {\cal A}_{12} & {\cal A}_{13} & \cdots \\
    {\cal A}_{21} & {\cal A}_{22} & {\cal A}_{23} & \cdots \\
\hline
    {\cal A}_{31} & {\cal A}_{32} & {\cal A}_{33} & \cdots \\
    \cdots &\cdots & \cdots& \cdots\\
  \end{array}
\right)\, .
\en
The singularities are contained in the upper left corner:
\eq
{\cal A}_{11}&=&-\frac{6}{\pi L}\,\frac{1}{1-\kappa^2}-\frac{1}{a}+\bar{\cal A}_{11}\, ,
\nonumber\\[2mm]
{\cal A}_{12}={\cal A}_{21}&=&-\frac{2\sqrt{6}}{\pi L}\,\frac{1}{1-\kappa^2}+\bar{\cal A}_{12}\, ,
\nonumber\\[2mm]
{\cal A}_{22}&=&-\frac{4}{\pi L}\,\frac{1}{1-\kappa^2}-\frac{1}{a}+\bar{\cal A}_{22}\, ,
\en
The rest of the matrix elements are regular. Note that, for further convenience, we have explicitly
singled out the leading term $1/a$ in the effective-range expansion as well.

In order to find the energy shift to leading order, it suffices to diagonalize
the singular part of the matrix. It is easily seen that this can be achieved through the
orthogonal transformation ${\cal A}\to \tilde{\cal O}{\cal A}\tilde{\cal O}^T$, where
\eq
\tilde{\cal O}=
\left(
  \begin{array}{c c | c c}
    c & s & 0 & \cdots \\
    -s & c & 0 & \cdots\\
    \hline
    0 & 0 & 1 & \cdots\\
    \cdots & \cdots & \cdots & \cdots\\
  \end{array}
\right)\, ,\quad\quad
c=\sqrt{\frac{3}{5}}\, ,\quad s=\sqrt{\frac{2}{5}}\, .
\en
Applying the transformation, we get
\eq\label{eq:diagonalization}
\tilde{\cal O}{\cal A}\tilde{\cal O}^T=
\left(
  \begin{array}{c c | c }
    X+\cdots & \cdots & \cdots \\
   \cdots & -\frac{1}{a}+\cdots & \cdots \\
    \hline
    \cdots & \cdots & \cdots  \\
  \end{array}
  \right)\, ,
\en
where we have introduced the notation
\eq
X=-\frac{10}{\pi L}\frac{1}{1-\kappa^2}-\frac{1}{a}\, ,
\en
and the ellipses stand for the regular terms at $\kappa^2\to 1$. It is now clear that, at lowest
order, the quantization condition reduces to $X=0$ that gives:
\eq\label{eq:leading-A1}
\kappa^2-1=\frac{10a}{\pi L}+O(L^{-2})\, .
\en
As seen from Eq.~(\ref{eq:diagonalization}), the $2\times 2$
symmetric matrix, containing only singular
terms, has only one non-zero eigenvalue and hence, there exists an unique solution
of the quantization condition\footnote{Note that this
  property may not hold in higher excited states, if a given irrep $\Gamma$ appears
  multiple times in the (generally reducible) representation of the group $O_h$
  in the vector space, spanned by the individual momenta of three particles.
  Then, a given unperturbed level may split.} in the vicinity of $\kappa^2=1$, given
by Eq.~(\ref{eq:leading-A1}). Unfortunately, at higher orders, the expressions become
quite cumbersome and a slightly different technique should be used. With a trial and
error method, we found it convenient to transform the matrix ${\cal A}$ by using a
different matrix ${\cal O}$, where
\eq
{\cal O}=
\left(
  \begin{array}{c c | c c}
    c & s & 0 & \cdots \\
    0 & 1 & 0 & \cdots\\
    \hline
    0 & 0 & 1 & \cdots\\
    \cdots & \cdots & \cdots & \cdots\\
  \end{array}
\right)\, .
\en
Note that ${\cal O}$ is not an orthogonal matrix.

The transformed matrix ${\cal O}{\cal A}{\cal O}^T$ takes the form
\eq
{\cal O}{\cal A}{\cal O}^T=
\left(
  \begin{array}{c c | c c}
    X+\bar {\cal A}_{11}' & sX+\bar{\cal A}_{12}' & {\cal A}_{13}' & \cdots \\
   sX+\bar{\cal A}_{21}' & s^2X-c^2/a+\bar{\cal A}_{22} & {\cal A}_{23} & \cdots\\
    \hline
    {\cal A}_{31}' & {\cal A}_{32} & {\cal A}_{33} &\cdots \\
    \cdots & \cdots &\cdots & \cdots \\
  \end{array}
  \right)\, ,
\en
where the ``primed'' matrix elements are given by:
 \eq
 \bar{\cal A}_{11}'&=&c^2\bar{\cal A}_{11}+2cs\bar{\cal A}_{12}+s^2\bar{\cal A}_{22}\, ,
 \nonumber\\[2mm]
 \bar{\cal A}_{12}'=\bar{\cal A}_{21}'&=&c\bar{\cal A}_{12}+s\bar{{\cal A}}_{22}\, ,
 \nonumber\\[2mm]
{\cal A}_{1i}'={\cal A}_{i1}'&=&c{\cal A}_{1i}+s{\cal A}_{2i}\, ,\quad\quad
i=2,\cdots .
\en
We now may use Eq.~(\ref{eq:series}) to obtain
\eq\label{eq:series1}
X+\bar {\cal A}_{11}'
=\sum_{k\neq 1}{\cal A}_{1k}'\tau_k{\cal A}_{k1}'
+\sum_{k,l\neq 1}{\cal A}_{1k}'\tau_kZ_{kl}\tau_l{\cal A}_{k1}'+\cdots\, .
\en
In order to single out all singular terms, we further define
\eq
{\cal M}_{kl}&=&Z_{kl}+\sum_{m\neq 1}Z_{km}\tau_m{\cal M}_{ml}\, ,\quad\quad
k,l=2,\cdots\, ,
\nonumber\\[2mm]
\hat {\cal M}_{kl}&=&Z_{kl}+\sum_{m\neq 1,2}Z_{km}\tau_m\hat {\cal M}_{ml}\, .
\en
>From the above equations, it is straightforward to obtain:
\eq
{\cal M}_{kl}=\hat{\cal M}_{kl}+\hat{\cal M}_{k2}\frac{1}{\tau_2^{-1}-\hat{\cal M}_{22}}\,
\hat{\cal M}_{2l}\, .
\en
The goal is achieved, because only the quantities $\tau_2^{-1},\hat{\cal M}_{22}$ contain singularities and everything else is free from singularities. Further, it can be checked that
the infinite
sum in Eq.~(\ref{eq:series1}) can be rewritten in terms of the matrix $\hat{\cal M}_{kl}$:
\eq
\sum_{k\neq 1}{\cal A}_{1k}'\tau_k{\cal A}_{k1}'
+\sum_{k,l\neq 1}{\cal A}_{1k}'\tau_kZ_{kl}\tau_l{\cal A}_{l1}'+\cdots
=\frac{({\cal A}_{12}'- c\bar\Delta_{12}-s\bar\Delta_{22})^2}
{\tau_2^{-1}-\hat{\cal M}_{22}}
+c^2\bar\Delta_{11}+2cs\bar\Delta_{12}+s^2\bar\Delta_{22}\, ,
\nonumber\\
\en
where the quantity
\eq
\bar\Delta_{kl}=\hat{\cal M}_{kl}-Z_{kl}
\en
does not contain singular terms.

Now, it is straightforward to obtain:
\eq
\tau_2^{-1}-\hat{\cal M}_{22}&=&s^2X-\frac{c^2}{a}+\bar{\cal A}_{22}-\bar\Delta_{22}\, ,
\nonumber\\[2mm]
{\cal A}_{12}'&=&sX+c\bar{\cal A}_{12}+s\bar{\cal A}_{22}\, .
\en
The quantization condition, Eq.~(\ref{eq:series1}), now takes  the form
\eq\label{eq:X0}
X
=\frac{(sX+cu_{12}+su_{22})^2}
{s^2X-\dfrac{c^2}{a}+u_{22}}
-(c^2u_{11}+2csu_{12}+s^2u_{22})\, ,\quad\quad
u_{kl}=\bar{\cal A}_{kl}-\bar\Delta_{kl}\, .
\en
Solving this equation with respect to $X$ finally gives:
\eq\label{eq:X1}
X=-\frac{(c^2u_{11}+2csu_{12}+s^2u_{22})-a(u_{11}u_{22}-u_{12}^2)}
{1-a(c^2u_{22}-2csu_{12}+s^2u_{11})}\, .
\en
At the end of this rather lengthy derivation, we have achieved our goal: all singular terms
are contained in the quantity $X$, whereas the quantities $\bar{\cal A}_{kl}$ and
$\bar\Delta_{kl}$ (and, hence, $u_{kl}$), are non-singular. The expansion for these quantities is given by
(cf. with Eq.~(\ref{eq:Delta})):
\eq\label{eq:Delta-A1}
\bar\Delta_{kl}=\frac{\bar\Delta_{kl}^{(0)}}{L^2}
+\frac{\bar\Delta_{kl}^{(1)}}{L^3}\,\ln\frac{mL}{2\pi}
+\frac{\bar\Delta_{kl}^{(2)}}{L^3}+\frac{\bar\Delta_{kl}^{(3)}}{L^2}\,(\kappa^2-1)+\cdots\, .
\en
Hence, the expansion of the quantity $\bar\Delta_{kl}$ starts at order $L^{-2}$. On the
other hand, the expansion of $\bar{\cal A}_{kl}$ can be read off
from Eqs.~(\ref{eq:tauL1}), (\ref{eq:tauL2}) and (\ref{eq:ZA1}) -- it starts
at order $L^{-1}$:
\eq\label{eq:barA}
\bar{\cal A}_{kl}=\biggl(\frac{a_{kl}^{(0)}}{L}+\frac{b_{kl}^{(0)}}{L^2}\biggr)
+\biggl(\frac{a_{kl}^{(1)}}{L}+\frac{b_{kl}^{(1)}}{L^2}\biggr)(1-\kappa^2)
+\frac{a_{kl}^{(2)}}{L}\,(1-\kappa^2)^2+\frac{c_{kl}}{L^3}\cdots
\en
Since $u_{kl}=O(L^{-1})$, the right-hand side of Eq.~(\ref{eq:X1}) can be expanded.
We retain only those terms that contribute to the energy shift at order $L^{-6}$: 
\eq\label{eq:A}
X&=&-(c^2u_{11}+2csu_{12}+s^2u_{22})
-a(cs(u_{22}-u_{11})+(c^2-s^2)u_{12})^2
\nonumber\\[2mm]
&-&a^2(cs(u_{22}-u_{11})+(c^2-s^2)u_{12})^2(c^2u_{22}-2csu_{12}+s^2u_{11})
+\cdots\, .
\en
As in case of the ground state, we solve the above equation iteratively. The explicit
expression for the energy shift at this order is given by
\eq
\kappa^2-1=\frac{h_0}{L}\biggl(1+\frac{h_1}{L}+
\frac{h_2}{L^2}+\frac{h_3}{L^3}\,\ln\frac{mL}{2\pi}
+\frac{h_4}{L^3}+\cdots\biggr)\, ,\quad\quad h_0=\frac{10a}{\pi}\, .
\label{eq:final-excited}
\en
Our aim is to determine the coefficients $h_1,h_2,h_3,h_4$. Substituting the above solution
into Eqs.~(\ref{eq:Delta-A1}) and (\ref{eq:A}) and recalling the definition of $u_{kl}$
from Eq.~(\ref{eq:X0}), we get
\eq
u_{kl}=\frac{u_{kl}^{(0)}}{L}+\frac{u_{kl}^{(1)}}{L^2}
+\frac{u_{kl}^{(2)}}{L^3}\,\ln\frac{mL}{2\pi}+\frac{u_{kl}^{(3)}}{L^3}+\cdots\, ,
\en
where
\eq
u_{kl}^{(0)}&=&a_{kl}^{(0)}\, ,
\nonumber\\[2mm]
u_{kl}^{(1)}&=&b_{kl}^{(0)}-a_{kl}^{(1)}h_0-\bar\Delta_{kl}^{(0)}\, ,
\nonumber\\[2mm]
u_{kl}^{(2)}&=&-\bar\Delta_{kl}^{(1)}\, ,
\nonumber\\[2mm]
u_{kl}^{(3)}&=&-b_{kl}^{(1)}h_0-a_{kl}^{(1)}h_0h_1+a_{kl}^{(2)}h_0^2+c_{kl}
-\bar\Delta_{kl}^{(2)}-\bar\Delta_{kl}^{(3)}h_0\, .
\en
Solving the quantization condition, we finally obtain:
\eq\label{eq:hi}
h_1&=&a\biggl(c^2u_{11}^{(0)}+2csu_{12}^{(0)}+s^2u_{22}^{(0)}\biggr)\, ,
\nonumber\\[2mm]
h_2&=&a\biggl(c^2u_{11}^{(1)}+2csu_{12}^{(1)}+s^2u_{22}^{(1)}\biggr)
\nonumber\\[2mm]
&+&a^2\biggl(c^2(u_{11}^{(0)})^2+s^2(u_{22}^{(0)})^2+(u_{12}^{(0)})^2
+2csu_{12}^{(0)}(u_{11}^{(0)}+u_{22}^{(0)})\biggr)\, ,
\nonumber\\[2mm]
h_3&=&a\biggl(c^2u_{11}^{(2)}+2csu_{12}^{(2)}+s^2u_{22}^{(2)}\biggr)\, ,
\nonumber
\en
\eq
h_4&=&a\biggl(c^2u_{11}^{(3)}+2csu_{12}^{(3)}+s^2u_{22}^{(3)}\biggr)
\nonumber\\[2mm]
&+&2a^2\biggl(c^2u_{11}^{(0)}u_{11}^{(1)}+s^2u_{22}^{(0)}u_{22}^{(1)}
+u_{12}^{(0)}u_{12}^{(1)}+csu_{12}^{(0)}(u_{11}^{(1)}+u_{22}^{(1)})
+csu_{12}^{(1)}(u_{11}^{(0)}+u_{22}^{(0)})\biggr)
\nonumber\\[2mm]
&+&a^3\biggl(2cs(u_{12}^{(0)})^3+(u_{12}^{(0)})^2((1+c^2)u_{11}^{(0)}
+(1+s^2)u_{22}^{(0)})
\nonumber\\[2mm]
&+&2csu_{12}^{(0)}((u_{11}^{(0)})^2+u_{11}^{(0)}u_{22}^{(0)}+(u_{22}^{(0)})^2)+(c^2(u_{11}^{(0)})^3+s^2(u_{22}^{(0)})^3)\biggr)\, .
\en
This solves the quantization condition at order $L^{-6}$. In order to complete the solution,
we should give explicit expressions for the coefficients $u_{kl}^{(\alpha)}$, entering
the above expressions.

\subsection{Expansion of $\bar{\cal A}_{kl}$}

The coefficients of the expansion of the quantities $\bar{\cal A}_{kl},~(k,l =1,2)$
can be directly
read off from Eqs.~(\ref{eq:tauL1}), (\ref{eq:tauL2}) and (\ref{eq:ZA1}). 
The result is given below:
\eq
a_{11}^{(0)}&=&-\frac{1}{\pi}\,{\cal I}_1+\frac{2}{\pi}\, ,\quad\quad
a_{11}^{(1)}=\frac{1}{\pi}\,{\cal J}_1+\frac{2}{\pi}\, ,\quad\quad
a_{11}^{(2)}=-\frac{1}{\pi}\,{\cal K}_1+\frac{2}{\pi}\, ,
\nonumber\\[2mm]
b_{11}^{(0)}&=&2\pi^2r\, ,\quad\quad
b_{11}^{(1)}=-2\pi^2r\, ,\quad\quad
c_{11}=-8\pi\frac{H_0(\Lambda)}{\Lambda^2}\, ,
\\[2mm]
a_{12}^{(0)}&=&
a_{12}^{(1)}
=a_{12}^{(2)}
=b_{12}^{(0)}
=b_{12}^{(1)}
=0\, ,\quad\quad
c_{12}=-8\sqrt{6}\pi\frac{H_0(\Lambda)}{\Lambda^2}\, ,
\\[2mm]
a_{22}^{(0)}&=&-\frac{1}{\pi}\,{\cal I}_2-\frac{9}{\pi}\, ,\quad\quad
a_{22}^{(1)}=\frac{1}{\pi}\,{\cal J}_2+\frac{17}{2\pi}\, ,\quad\quad
a_{22}^{(2)}=-\frac{1}{\pi}\,{\cal K}_2-\frac{33}{4\pi}\, ,
\nonumber\\[2mm]
b_{22}^{(0)}&=&\frac{1}{2}\,\pi^2r\, ,\quad\quad
b_{22}^{(1)}=-2\pi^2r\, ,\quad\quad
c_{22}=-48\pi\frac{H_0(\Lambda)}{\Lambda^2}\, .
\en

\subsection{Expansion of $\bar\Delta_{kl}$}

In order to obtain the expansion of $\bar\Delta_{kl}$, we have no consider the quantity
\eq\label{eq:Delta-def}
\Delta({\bf p},{\bf q};E)=\frac{1}{L^3}\,{\sum_{\bf k}}'
Z({\bf p},{\bf k};E)\tau_L({\bf k};E)Z({\bf k},{\bf q};E)+\cdots\, ,
\en
where ${\bf p}=\dfrac{2\pi}{L}\,{\bf i}$ and ${\bf q}=\dfrac{2\pi}{L}\,{\bf j}$
belong either to shell $s=1$ or $s=2$, and the prime over
the summation indicates that ${\bf k}^2=\left(\dfrac{2\pi}{L}\right)^2{\bf n}^2>\left(\dfrac{2\pi}{L}\right)^2$ (the vector ${\bf k}$ belongs to the shell $r>2$).

The expansion of the above quantity in the limit of the large $L$ proceeds in the same
way as for the ground state:
\eq\label{eq:Delta-ex}
&&
L^{-3}\Delta({\bf p},{\bf q};E)=\frac{\Delta^{(0)}({\bf p},{\bf q})}{L^2}+
\frac{\Delta^{(1)}({\bf p},{\bf q})}{L^3}\,\ln\frac{mL}{2\pi}
+\frac{\Delta^{(2)}({\bf p},{\bf q})}{L^3}+(\kappa^2-1)
\frac{\Delta^{(3)}({\bf p},{\bf q})}{L^2}+O(L^{-4})\, .
\nonumber\\
&&
\en
where
\eq
\Delta^{(0)}({\bf p},{\bf q})&=&-\frac{a{\cal J}({\bf i},{\bf j})}{2\pi^3}\, ,
\nonumber\\[2mm]
\Delta^{(1)}({\bf p},{\bf q})&=&-\frac{2\sqrt{3}a^2}{\pi}+\frac{8a^2}{3}\, ,
\nonumber\\[2mm]
\frac{H_0(\Lambda)}{\Lambda^2}+\Delta^{(2)}({\bf p},{\bf q})&=&\hat{\cal M}+\frac{a^2}{(2\pi)^3}\,{\cal B}({\bf i},{\bf j})\, ,
\nonumber\\[2mm]
\Delta^{(3)}({\bf p},{\bf q})&=&-\frac{a}{\pi^3}\,{\cal K}({\bf i},{\bf j})\, ,
\en
the integer vectors ${\bf i},{\bf j}$ are given by
${\bf i}=\dfrac{L{\bf p}}{2\pi}\,,~
      {\bf j}=\dfrac{L{\bf q}}{2\pi}\,,$
and
\eq
{\cal J}({\bf i},{\bf j})
&=&\sum_{|{\bf n}|\neq 0,1}^\infty
\frac{1}{{\bf i}^2+{\bf i}{\bf n}+{\bf n}^2-1}\,
\frac{1}{{\bf n}^2+{\bf n}{\bf j}+{\bf j}^2-1}\, ,
\nonumber\\[2mm]
{\cal K}({\bf i},{\bf j})
&=&\frac{1}{2}\,\sum_{|{\bf n}|\neq 0,1}^\infty
\biggl(\frac{1}{({\bf i}^2+{\bf i}{\bf n}+{\bf n}^2-1)^2}\,
\frac{1}{{\bf n}^2+{\bf n}{\bf j}+{\bf j}^2-1}
\nonumber\\[2mm]
&+&\frac{1}{{\bf i}^2+{\bf i}{\bf n}+{\bf n}^2-1}\,
\frac{1}{({\bf n}^2+{\bf n}{\bf j}+{\bf j}^2-1)^2}\biggr)\, ,
\nonumber\\[2mm]
{\cal B}({\bf i},{\bf j})
&=&
  -16\pi^2\sqrt{3}
\biggl(\frac{1}{2}\,(1-\ln 3)+\frac{\sqrt{3}\pi}{18}\biggr)
-4\pi\sqrt{3}{\cal L}({\bf i},{\bf j})
+\frac{4}{\pi}\,{\cal G}({\bf i},{\bf j})
+\frac{8}{\pi}\,
  (2{\cal Q}({\bf i},{\bf j})-{\cal I}_0)\, .
\nonumber \\[2mm]
{\cal L}({\bf i},{\bf j})
&=&\lim_{N\to\infty}\biggl(\sum_{|{\bf n}|\neq 0,1}^N
\frac{\sqrt{\dfrac{3}{4}\,{\bf n}^2-1}}{({\bf i}^2+{\bf i}{\bf n}+{\bf n}^2-1)
({\bf n}^2+{\bf n}{\bf j}+{\bf j}^2-1)}-4\pi\ln N\biggr)\, ,
\nonumber \\[2mm]
{\cal G}({\bf i},{\bf j})
&=&\sum_{|{\bf n}|\neq 0,1}^\infty
\frac{{\cal I}(\bf n)}{({\bf i}^2+{\bf i}{\bf n}+{\bf n}^2-1)
({\bf n}^2+{\bf n}{\bf j}+{\bf j}^2-1)}
\nonumber \\[2mm]
{\cal Q}({\bf i},{\bf j})
&=&\lim_{N\to\infty}\biggl(
\frac{1}{2}\,\sum_{|{\bf n}|\neq 0,1}^N\sum_{|{\bf m}|\neq 0,1}^N
\frac{1}{{\bf i}^2+{\bf i}{\bf n}+{\bf n}^2-1}\,
\frac{1}{{\bf n}^2+{\bf n}{\bf m}+{\bf m}^2-1}\,
\frac{1}{{\bf m}^2+{\bf m}{\bf j}+{\bf j}^2-1}
\nonumber\\[2mm]
&-&\frac{4\pi^4}{3}\,\ln N\biggr)\, .
\en
Note that the quantities, defined in the infinite volume (the threshold amplitude
$\hat {\cal M}$, the coefficient in front of the logarithm and the integral ${\cal I}_0$)
are the same as in the case of the ground-state shift.

Recalling now the definition
\eq
\bar\Delta_{kl}=\frac{8\pi\sqrt{\theta(k)\theta(l)}}{GL^3}\,
\sum_{g\in O_h}\Delta(g{\bf p},{\bf q},E)\, ,
\en
we get:
\eq
\bar\Delta_{kl}^{(0)}&=&-\frac{4a}{\pi^2}\,{\cal J}_{kl}\, , 
\nonumber\\[2mm]
\bar\Delta_{kl}^{(1)}&=&\biggl(-16\sqrt{3}+\frac{64\pi}{3}\biggr)a^2\sqrt{\theta(k)\theta(l)}\, ,
\nonumber\\[2mm]
\bar\Delta_{kl}^{(2)}-c_{kl}&=&8\pi\hat{\cal M}\sqrt{\theta(k)\theta(l)}+\frac{a^2}{\pi^2}\,
{\cal B}_{kl}\, ,
\nonumber\\[2mm]
\bar\Delta_{kl}^{(3)}&=&-\frac{8a}{\pi^2}\,{\cal K}_{kl}\, ,
\en
where
\eq
{\cal B}_{kl}=\biggl[-16\pi^2\sqrt{3}
\biggl(\frac{1}{2}\,(1-\ln 3)+\frac{\sqrt{3}\pi}{18}\biggr)
-\frac{8}{\pi}\,{\cal I}_0\biggr]\sqrt{\theta(k)\theta(l)}
-4\pi\sqrt{3}{\cal L}_{kl}
+\frac{4}{\pi}\,{\cal G}_{kl}
+\frac{16}{\pi}\,{\cal Q}_{kl}\, .
\en
In the above equations, $\theta(1)=1$, $\theta(2)=6$, and the quantities
${\cal J}_{kl},~ {\cal K}_{kl},~ {\cal L}_{kl},~ {\cal G}_{kl},~ {\cal Q}_{kl}$
are defined in Appendix~\ref{app:sums}.

Our derivation of the energy shift of the first excited level in the irrep $A_1^+$ is now complete -- all quantities, contributing to $h_1,h_2,h_3,h_4$ from Eq.~(\ref{eq:hi}) are explicitly defined. Next, we shall demonstrate, how an analogous formula can be obtained in the irrep $E^+$.

\subsection{The quantization condition in the irrep $E^+$}
  
The irrep $E^+$ is two-dimensional. Thus, in the quantization condition
(\ref{eq:quant-cond}), the labels $\sigma,\rho$ now run from 1 to 2. 
Let
us start from shell 1. Projecting\footnote{The matrices of different irreps
of the octahedral group and the little groups thereof can be found
in Refs.~\cite{Bernard:2008ax,Gockeler:2012yj,Romero-Lopez:2018zyy}.}
the driving term $Z$ onto the irrep  $E^+$, we obtain that, for an arbitrary 
$s$,
\eq\label{eq:Z1}
Z_{\sigma\rho}^{E^+}(1,s)=Z_{\sigma\rho}^{E^+}(s,1)=0\, .
\en
At a first glance, this leads to the controversy since, as seen from
Eq.~(\ref{eq:quant-cond}), one of the roots decouples from the rest and 
obeys the equation
\eq
\tau_L^{-1}(s=1,E)=0\, .
\en
This root would correspond to the pure two-body level, without an interaction with 
the spectator. It is, however, easy to show that we are dealing here with a spurious solution. To
this end, one has to return one step back and consider the system of linear
equations, which yield the quantization condition, prior to projecting onto
the different irreps. Taking into account Eq.~(\ref{eq:Apq}), this system
is written down as
\eq
\tau_L^{-1}({\bf p};E)f({\bf p})-\frac{1}{L^3}\,\sum_{\bf q}Z({\bf p},{\bf q};E)f({\bf q})=0\, ,
\en
see also Ref.~\cite{Doring:2018xxx}. Here, $f({\bf p})$ is a solution.
This system has non-trivial solutions, if and only if the determinant vanishes,
that leads to the quantization condition already considered above.

Projecting now this system of equations onto the irrep $E^+$, it is seen that
the projection of the vector $f({\bf p})$ on the first shell, corresponding
to ${\bf p}=0$, vanishes identically due to symmetry reasons:
\eq
\sum_{g\in O_h} (T_{\sigma\rho}^{E^+}(g))^*f({\bf 0})=0\, ,
\en
and, hence,
the first two lines $s=1,~\sigma=1,2$
of the projected equations, which correspond to the projection
onto the first shell, turn into a trivial identity $0=0$ and should
be deleted (we remind the reader
that the projection of the driving term $Z$ onto the first shell also
vanishes, see Eq.~(\ref{eq:Z1})). Thus, the spurious solution disappears.
This simple example shows a pattern, suggesting how to avoid the spurious solutions
in general: if a projection $f_{\rho\sigma}^\Gamma(s)$ of a generic 
vector $f({\bf p})$
identically vanishes for a given $s$ and $\rho$ (the second index $\sigma$
is fixed, see Ref.~\cite{Doring:2018xxx} for details), one has to delete
the corresponding rows/columns in the quantization condition from
the beginning. Note also that deleting the entries corresponding to the
first shell, makes the solution of the quantization condition easier --
unlike the case with the irrep $A_1^+$, now there is no need of the 
special treatment of the singular terms in the matrix ${\cal A}$.
 
Now, let us turn to the shell $s=2$, corresponding to $|{\bf p}|=2\pi/L$ and
choose the reference momentum as ${\bf p}_0=2\pi/L\,(0,0,1)$. In total, there
are six different momenta in the second shell, which are obtained from the reference momentum
by acting the group elements belonging to the set ${\cal S}=\{g_1=\mathbb{I},g_2,g_3,I,Ig_2,Ig_3\}$ ($\mathbb{I}$ is the identity matrix and $I$ denotes
the inversion of all components of the vector; 
the numeration is the same as, e.g., in Table
B.1 of Ref.~\cite{Romero-Lopez:2018zyy}). 
Namely, $g_2{\bf p}_0={\bf p}_0'=2\pi/L\,(0,1,0)$ and
$g_3{\bf p}_0={\bf p}_0''=2\pi/L\,(1,0,0)$.
Further, let $\hat g$ denote
the elements of a subgroup in $O_h$ that leave the reference
vector ${\bf p}_0$ invariant. This subgroup $\hat {\cal G}$ consists of the following
elements: $\hat g=g_1,g_{14},g_{15},g_{24},Ig_{18},Ig_{19},Ig_{22},Ig_{23}$.
Consequently,
\eq
Z^{E^+}_{\sigma\rho}(2,s)&=&\sum_{g'\in{\cal S}}\sum_{\hat g\in\hat{\cal G}}
(T_{\rho\sigma}^{E^+}(g'\hat g))^*Z(g'\hat g{\bf p}_0,{\bf q}_0;E)
\nonumber\\[2mm]
&=&\sum_{g'\in{\cal S}}(T_{\rho\lambda}^{E^+}(g'))^*Z(g'{\bf p}_0,{\bf q}_0;E)
\biggl\{\sum_{\hat g\in\hat{\cal G}}(T_{\lambda\sigma}^{E^+}(\hat g))^*\biggr\}
=\begin{pmatrix}
a(s) & 0\\
b(s) & 0
\end{pmatrix}_{\rho\sigma}\, ,
\en
where ${\bf q}_0$ denotes the reference momentum for the shell $s$ and
\eq
a(s)&=&8\biggl(Z({\bf p}_0,{\bf q}_0;E)+Z(-{\bf p}_0,{\bf q}_0;E)
-\frac{1}{2}\,Z({\bf p}_0',{\bf q}_0;E)-\frac{1}{2}\,Z(-{\bf p}_0',{\bf q}_0;E)
\nonumber\\[2mm]
&-&\frac{1}{2}\,Z({\bf p}_0'',{\bf q}_0;E)-\frac{1}{2}\,Z(-{\bf p}_0'',{\bf q}_0;E)\biggr)\, ,
\nonumber\\[2mm]
b(s)&=&-4\sqrt{3}\biggl(Z({\bf p}_0',{\bf q}_0;E)+Z(-{\bf p}_0',{\bf q}_0;E)
-Z({\bf p}_0'',{\bf q}_0;E)-Z(-{\bf p}_0'',{\bf q}_0;E)\biggr)\, .
\en
In particular, $b(s)=0$ for the shell $s=2$.

Now, it is straightforward to convince oneself that the non-diagonal matrix
elements of the matrix ${\cal A}^{2\sigma}_{2s}$ and ${\cal A}^{\sigma 2}_{s2}$
vanish for $\sigma=1,2$. Using the same method as for the shell $s=1$, it can
be shown that this row/column in the matrix ${\cal A}$ should be deleted, in
order to avoid spurious solutions. Consequently, the quantization
condition is rewritten in the following form\footnote{In the matrix ${\cal A}^{\sigma\rho}_{rs}$ the lower indices, as before, label the shells, whereas the upper indices $\sigma,\rho=1,2$ are used to label the basis vectors of the irrep
$E^+$. In order to avoid clutter of various indices, we further omit the
label $E^+$ from all expressions.}
\eq
{\cal A}^{11}_{22}=\Delta^{11}_{22}\, ,
\en
where
\eq
\Delta^{11}_{22}=\frac{48\pi}{GL^3}\sum_{g\in O_h}(T_{11}^{E^+}(g))^*
\Delta(g{\bf p}_0,{\bf p}_0;E)\, ,
\en
where ${\bf p}_0$ is the reference momentum for the shell $s=2$ and
the quantity $\Delta({\bf p},{\bf p};E)$ is defined in Eq.~(\ref{eq:Delta-def}).
Further,
\eq
{\cal A}^{11}_{22}=-\frac{1}{a}-\frac{4}{\pi L}\,\frac{1}{1-\kappa^2}+
\bar {\cal A}^{11}_{22}\, ,
\en
the expansion of the quantity $\bar {\cal A}^{11}_{22}$ is similar but not
identical
to Eq.~(\ref{eq:barA}) (note the absence of the term with $c_{22}$):
\eq
\bar{\cal A}^{11}_{22}=\biggl(\frac{{a_{22}^{(0)}}'}{L}+\frac{{b_{22}^{(0)}}'}{L^2}\biggr)
+\biggl(\frac{{a_{22}^{(1)}}'}{L}+\frac{{b_{22}^{(1)}}'}{L^2}\biggr)(1-\kappa^2)
+\frac{{a_{22}^{(2)}}'}{L}\,(1-\kappa^2)^2+\cdots\, .
\en
where
\eq 
{a_{22}^{(0)}}'&=&-\frac{1}{\pi}\,{\cal I}_2+\frac{3}{\pi}\, ,\quad\quad
{a_{22}^{(1)}}'=\frac{1}{\pi}\,{\cal J}_2-\frac{7}{2\pi}\, ,\quad\quad
{a_{22}^{(2)}}'=-\frac{1}{\pi}\,{\cal K}_2+\frac{15}{4\pi}\, ,
\nonumber\\[2mm]
{b_{22}^{(0)}}'&=&\frac{1}{2}\,\pi^2r\, ,\quad\quad
{b_{22}^{(1)}}'=-2\pi^2r\, .
\en
Further, the expansion of the quantity $\Delta^{11}_{22}$ is given by,
cf. with Eq.~(\ref{eq:Delta-A1}):
\eq
\Delta^{11}_{22}=\frac{{\Delta_{22}^{(0)}}'}{L^2}
+\frac{{\Delta_{22}^{(2)}}'}{L^3}+\frac{{\Delta_{22}^{(3)}}'}{L^2}\,(\kappa^2-1)+\cdots\, ,
\en
where
\eq
{\Delta_{22}^{(0)}}'&=&-\frac{4a}{\pi^2}\,{\cal J}'_{22}\, ,
\nonumber\\[2mm]
{\Delta_{22}^{(2)}}'&=&\frac{a^2}{\pi^2}\,\biggl\{-4\pi\sqrt{3}{\cal L}'_{22}
+\frac{4}{\pi}\,{\cal G}'_{22}+\frac{16}{\pi}\,{\cal Q}'_{22}\biggr\}\, ,
\nonumber\\[2mm]
{\Delta_{22}^{(3)}}'&=&-\frac{8a}{\pi^2}\,{\cal K}'_{22}\, .
\en
Note again the absence of the logarithmic term and the threshold amplitude,
as compared to the case of the irrep $A_1^+$.
The numerical constants in the above equation are given in Appendix~\ref{app:sums}.

Finally, the expansion of the energy shift is given by
\eq
\kappa^2-1=\frac{h_0'}{L}\,\biggl(1+\frac{h_1'}{L}+\frac{h_2'}{L^2}+\frac{h_4'}
      {L^3}\biggr)\, ,
      \label{eq:final-excited-E}
\en
where
 \eq
h_0'&=&\frac{4a}{\pi}\, ,
\nonumber\\[2mm]
h_1'&=&a{u_{22}^{(0)}}'\, ,
\nonumber\\[2mm]
h_2'&=&\bigl(a{u_{22}^{(0)}}'\bigr)^2
+a{u_{22}^{(1)}}'\, ,
\nonumber\\[2mm]
h_4'&=&\bigl(a{u_{22}^{(0)}}'\bigr)^3
+2a^2{a_{22}^{(0)}}'{u_{22}^{(1)}}'
+a{u_{22}^{(3)}}'\, ,
\en
and
\eq
{u_{22}^{(0)}}'&=&{a_{22}^{(0)}}'\, ,
\nonumber\\[2mm]
{u_{22}^{(1)}}'&=&{b_{22}^{(0)}}'-{a_{22}^{(1)}}'h_0'-{\Delta_{22}^{(0)}}'\, ,
\nonumber\\[2mm]
{u_{22}^{(3)}}'&=&-{b_{22}^{(1)}}'h_0'-{a_{22}^{(1)}}'h_0'h_1'
+{a_{22}^{(2)}}'(h_0')^2-{\Delta_{22}^{(2)}}'-{\Delta_{22}^{(3)}}'h_0'\, .
\en
This is the final expression for the energy shift at order $L^{-6}$. As one 
sees, at this order it contains neither the threshold amplitude, nor
the logarithmic term. The reason for this is the symmetry: both terms do not
depend on the direction. Averaging them over the octahedral group yields zero
in all irreps except $A_1^+$. A particle-dimer interaction in higher partial
waves would potentially contribute here but, according to the power counting,
 this contribution comes at higher order in $1/L$.

\section{Comparison to some earlier work}
\label{sec:comparison}

\subsection{Comparison with Hansen and Sharpe}

First of all, it should be pointed out that the framework, used in the present paper, is
non-relativistic. Therefore, the relativistic corrections, which are present in the expression
from Refs.~\cite{Hansen:2015zta,Hansen:2016fzj,Sharpe:2017jej} at order $L^{-6}$,
cannot be reproduced. The work on the relativistic
version of our approach is in progress, and the results will be reported elsewhere.

Further, in order to carry out the comparison, we have to relate our threshold amplitude
$\hat {\cal M}$ with the amplitude $M_{3,thr}$, which appears in the above papers.
We remind the reader that our threshold amplitude $\hat {\cal M}$ is defined as follows:
consider the particle-dimer scattering amplitude ${\cal M}({\bf p},{\bf q};E)$ at
${\bf p}={\bf q}=0$ and $-mE=\varepsilon^2\to 0$. This amplitude can be written
in the following form:
\eq\label{eq:012}
{\cal M}=I_0+I_1+I_2+\bar{\cal M}\, ,
\en
where $I_0$ is the tree-level exchange term and $I_1,I_2$ denote the first and the second
iteration of $I_0$. All the rest, which is denoted by $\bar{\cal M}$, is regular, as
$\varepsilon\to 0$, whereas $I_0,I_1,I_2$ contain the singularities
$\varepsilon^{-2},\varepsilon^{-1}$ and $\ln\dfrac{\varepsilon}{m}$.
It can be shown that
\eq
I_0+I_1+I_2=\frac{4a}{\pi\Lambda}-\frac{2a^2}{\pi}\,\ln\frac{\Lambda}{m}\,
\biggl(\sqrt{3}-\frac{4\pi}{3}\biggr)-\frac{a^2}{(2\pi)^3}\,\delta+\mbox{singular terms in }\varepsilon\, ,
\en
where $\delta$ is given in Eq.~(\ref{eq:delta}).
The threshold amplitude  $\hat {\cal M}$ is
obtained from ${\cal M}$ by deleting these singularities. Hence,
\eq
\hat{\cal M}=\frac{4a}{\pi\Lambda}-\frac{2a^2}{\pi}\,\ln\frac{\Lambda}{m}\,
\biggl(\sqrt{3}-\frac{4\pi}{3}\biggr)-\frac{a^2}{(2\pi)^3}\,\delta+\bar{\cal M}\, .
\en
On the other hand, following Refs.~\cite{Hansen:2015zta,Hansen:2016fzj,Sharpe:2017jej},
one may introduce a different splitting of the same amplitude:
\eq
{\cal M}=I_0^H+I_1^H+I_2^H+\bar{\cal M}^H\, ,
\en
where the superscript $H$ indicates that the ultraviolet regularization, instead
of the sharp cutoff $\Lambda$, is carried out by using the smooth cutoff function
$H$, introduced in these papers.
The prescription, used in Refs.~\cite{Hansen:2015zta,Hansen:2016fzj,Sharpe:2017jej},
consists in dropping the singular pieces $I_0^H,I_1^H,I_2^H$ altogether. The
particle-dimer threshold amplitude within this prescription is equal merely to
$\bar{\cal M}^H$. The three particle threshold amplitude ${\cal M}_3$ is,
modulo overall normalization, which will be considered later,
proportional to $\bar{\cal M}^H$.

>From these definitions, it is straightforward to obtain a linear relation between
$\hat{\cal M}$ and $\bar{\cal M}^H$:
\eq
\bar{\cal M}^H=\biggl(\hat{\cal M}+\frac{a^2}{(2\pi)^3}\,\delta\biggr)
+\Delta S'+\Delta S''+\Delta S'''\, ,
\en
where
\eq\label{eq:DeltaS}
\Delta S'&=&-8\pi a\int^\Lambda\frac{d^3{\bf k}}{(2\pi)^3}\,\frac{1}{{\bf k}^4}
-\frac{4a}{\pi\Lambda}
+8\pi a\int\frac{d^3{\bf k}}{(2\pi)^3}\,\frac{H^2\left(\dfrac{k}{m}\right)}{{\bf k}^4}\, ,
\nonumber\\[2mm]
\Delta S''&=&-4\sqrt{3}\pi a^2\int^\Lambda\frac{d^3{\bf k}}{(2\pi)^3}\,\frac{1}{|{\bf k}|^3}
+\frac{2a^2\sqrt{3}}{\pi}\,\ln\frac{\Lambda}{m}
  +4\sqrt{3}\pi a^2\int\frac{d^3{\bf k}}{(2\pi)^3}\,
  \frac{H^3\left(\dfrac{k}{m}\right)}{|{\bf k}|^3}\, ,
  \nonumber\\[2mm]
 \Delta S'''&=& 64\pi^2a^2
  \int^\Lambda\frac{d^3{\bf k}}{(2\pi)^3}\int^\Lambda\frac{d^3{\bf q}}{(2\pi)^3}\,
  \frac{1}{{\bf k}^2({\bf k}^2+{\bf k}{\bf q}+{\bf q}^2){\bf q}^2}
  -\frac{8a^2}{3}\ln\frac{\Lambda}{m}
\nonumber\\[2mm]
  &-&64\pi^2a^2
  \int\frac{d^3{\bf k}}{(2\pi)^3}\int\frac{d^3{\bf g}}{(2\pi)^3}\,
  \frac{H^2\left(\dfrac{k}{m}\right)H^2\left(\dfrac{q}{m}\right)}{{\bf k}^2({\bf k}^2+{\bf k}{\bf q}+{\bf q}^2){\bf q}^2}\, .
  \en
  Note that, in the above integrals, there is no need of infrared regularization anymore.

  In order to complete the comparison, one has next to relate various infinite sums
  over integer numbers, which appear at order $L^{-6}$ in
  Refs.~\cite{Hansen:2015zta,Hansen:2016fzj,Sharpe:2017jej} and in the present paper.
  In the above papers, the ultraviolet regularization is again carried out by introducing
  the smooth cutoff function $H$. Thus, the difference emerges only at high momenta
  due to the different ultraviolet cutoff and, consequently, one may safely perform the
  limit $L\to\infty$ in the differences, replacing the summation by integration. As a result,
  one obtains:
  \eq\label{eq:F345}
  {\cal C}_F&=&-4{\cal G}\, ,
  \nonumber\\[2mm]
  \Delta  S'&=&\frac{8\pi a}{m}\,{\cal C}_3\, ,
  \nonumber\\[2mm]
  \Delta  S''&=&\frac{4\pi a^2\sqrt{3}}{(2\pi)^3}\,
  \biggl( \frac{1}{4\pi^2\sqrt{3}}\,{\cal C}_4-{\cal L}\biggr)\, ,
  \nonumber\\[2mm]
 \Delta  S'''&=&\frac{a^2}{8\pi^4}\,{\cal C}_5+\frac{2a^2}{\pi^4}\,{\cal Q}\, ,
\en
where ${\cal C}_F,{\cal C}_3,{\cal C}_4,{\cal C}_5$ are the numerical constants, appearing in Refs.~\cite{Hansen:2015zta,Hansen:2016fzj,Sharpe:2017jej}. Taking into account
Eqs.~(\ref{eq:DeltaS}) and (\ref{eq:F345}), it is seen that all $H$-dependent terms
in the energy shift expression from Refs.~\cite{Hansen:2015zta,Hansen:2016fzj,Sharpe:2017jej} cancel (as they indeed should), and this expression algebraically coincides with our expression up to the replacement $\hat{\cal M}+\dfrac{a^2}{(2\pi)^3}\,\delta\to \bar{\cal M}^H$.

What remains to be fixed is the normalization coefficient in the relation
of $\bar{\cal M}^H$ and $M_ {3,thr}$. Most easily this can be done by comparing the
normalization of the tree-level exchange term $I_0$ in the present paper and in 
Refs.~\cite{Hansen:2015zta,Hansen:2016fzj,Sharpe:2017jej}. The result reads
\eq
9a^2\bar {\cal M}^H=\frac{1}{128\pi^2(2m)^2}\,M_{3,thr}\, ,
\en
where the combinatorial factor 9 emerges, when one relates the particle-dimer scattering
amplitude to the three-particle amplitude, with all permutations of particles in the initial
and final states. Using this relation, it is then seen that the energy shift from Refs.~\cite{Hansen:2015zta,Hansen:2016fzj,Sharpe:2017jej} exactly reproduces our result.

\subsection{Comparison with Beane, Detmold and Savage}

In order to compare our result with Ref.~\cite{Beane}, a perturbative expansion of
the quantity $\hat{\cal M}$ should be carried out. To this end, one has to relate the
particle-dimer coupling $\dfrac{H_0(\Lambda)}{\Lambda^2}$, which emerges in our
expressions, to the three-body coupling constant $\eta_3^r(\mu)$ from Ref.~\cite{Beane},
where the ultraviolet divergence is removed by using dimensional regularization, and
$\mu$ is the scale of the dimensional regularization. If there are no derivative two-body
couplings, that lead to the non-zero effective range, these two couplings at tree level
are merely proportional to each other. However, if the derivative terms are present, the
relation is more complicated, and can be figured out in the following manner.
Consider the particle-dimer Lagrangian\footnote{Here, for convenience, we use a dimer formulation that is different from one described in Ref.~\cite{pang2}. Namely, here the effective-range term is included in the kinetic term of the dimer and not in the derivative particle-dimer vertex. Both formulations are equivalent, leading to the same $S$-matrix.}
\eq
{\cal L}=\psi^\dagger\biggl(i\partial_0-\frac{\nabla^2}{2m}\biggr)\psi
+d^\dagger\biggl(i\partial_0+\frac{\nabla^2}{4m}+\Delta\biggr)d
-\frac{g}{\sqrt{2}}\,(d^\dagger\psi\psi+\mbox{h.c.})+hd^\dagger d\psi^\dagger\psi+\cdots\, ,
\en
where $\psi$ and $d$ stand for the non-relativistic particle and dimer fields, respectively.
The quantity $h$ is proportional to the particle-dimer coupling
$h=2mg^2\dfrac{H_0(\Lambda)}{\Lambda^2}$.

Let us now integrate out the dimer field $d$ that leads to the three-particle Lagrangian.
At tree level, one may set the emerging determinant to unity. Expanding the obtained
Lagrangian in powers of $\dfrac{1}{\Delta}$, we get
\eq
{\cal L}=\psi^\dagger\biggl(i\partial_0-\frac{\nabla^2}{2m}\biggr)\psi
-\frac{g^2}{2\Delta}\,\psi^\dagger\psi^\dagger\biggl\{1-\frac{1}{\Delta}\,
\biggl(i\partial_0+\frac{\nabla^2}{4m}-h\psi^\dagger\psi\biggr)+\cdots\biggr\}\psi\psi\, .
\en
Using the equations of motion
\eq
\biggl(i\partial_0+\frac{\nabla^2}{2m}-\frac{g^2}{\Delta}\,\psi^\dagger\psi\biggr)\psi=0
\en
in the last term, one can rewrite the Lagrangian in the following form
\eq
{\cal L}&=&\psi^\dagger\biggl(i\partial_0-\frac{\nabla^2}{2m}\biggr)\psi
-\frac{g^2}{2\Delta}\,(\psi^\dagger\psi)^2
-\frac{g^2}{4m\Delta^2}\,(\psi^\dagger\psi^\dagger\psi\nabla^2\psi+\mbox{h.c.})
\nonumber\\[2mm]
&+&\biggl(\frac{g^2h}{2\Delta^2}+\frac{g^4}{\Delta^3}\biggr)(\psi^\dagger\psi)^3
+\frac{g^2}{8m\Delta^2}\psi^\dagger\psi^\dagger\nabla^2(\psi\psi)+\cdots\, .
\en
>From this, one may read off the matching conditions of the couplings:
\eq
\frac{g^2}{\Delta}=\frac{4\pi a}{m}\, ,\quad
\frac{1}{m\Delta}=-\frac{1}{2}\, ar\, ,\quad
\eta_3=-6\biggl(\frac{g^2h}{2\Delta^2}+\frac{g^4}{\Delta^3}\biggr)
=-6m\biggl(\frac{4\pi a}{m}\biggr)^2\biggl(\frac{H_0(\Lambda)}{\Lambda^2}-\frac{1}{2}\,ar\biggr)\, .
\en
Here, $\eta_3$ is the six-particle coupling from Ref.~\cite{Beane}.

Now, we are in a position to discuss a seeming disagreement between our result
(which, as mentioned, agrees with Refs.~\cite{Hansen:2015zta,Hansen:2016fzj,Sharpe:2017jej}) and
the result of Ref.~\cite{Beane}. Namely, the effective-range term
equals to $\dfrac{6\pi a^2r}{L^3}$ in our approach vs $\dfrac{2\pi a^2r}{L^3}$
in Ref.~\cite{Beane}. The solution of the puzzle is as follows:
it is shown that the matching of the couplings $\dfrac{H_0(\Lambda)}{\Lambda^2}$
and $\eta_3$ at tree level contains an additive term $-\dfrac{1}{2}\,ar$ with a fixed
coefficient, which exactly cancels this difference and makes the results
consistent.\footnote{Note that such an effective range-dependent shift
  in the three-body coupling $H_0$ also appears in the infinite volume
  when the leading effective range correction is included
  \cite{Hammer:2001gh,Ji:2011qg}. The coefficient of this shift contains
  no free parameter and is fully determined by leading-order information.}

The calculation of the three-body energy shift has been carried out in Ref.~\cite{Tan}
as well by using a different technique. Here, we do not carry out a detailed comparison with
Ref.~\cite{Tan}, limiting ourselves to a brief comment. As the author of Ref.~\cite{Tan}
suggests, the results of Refs.~\cite{Beane} and \cite{Tan} are identical, if the three-body
couplings in both approaches are matched appropriately. In particular, the matching
condition involves an additive contribution, proportional to the effective range
$r$ (see, e.g., Eq.~(115) of Ref.~\cite{Tan}). We would like to stress here again that
the coefficient of the effective-range term at lowest order is fixed. Consequently, in
order to demonstrate the equivalence of the two approaches, it still remains to be shown
that one gets the right coefficient in the matching condition -- in analogy to the discussion above.

In order to complete the comparison of our result with Ref.~\cite{Beane}, we still have to calculate the
singular pieces of the threshold amplitude in the dimensional regularization which is used
in Ref.~\cite{Beane} (we are using the cutoff regularization throughout this paper).
We start with the contribution to the particle-dimer
amplitude, which corresponds to
one iteration of the driving term $Z$ (an infinite-volume analog of
$S_1$ considered in section~\ref{sec:one-iteration}).
Only the infrared-singular piece is of interest here (the rest is not shown explicitly).
Giving the energy a negative infinitesimal shift
$-mE=\varepsilon^2$ and writing down the dimer propagator as a sum of $0,1,2\ldots$
bubbles, this particular contribution is given by
\eq
S_1^d&=&-8\pi a\int\frac{d^d{\bf k}}{(2\pi)^d}\,\frac{1}{({\bf k}^2+\varepsilon^2)^2}
\nonumber\\[2mm]
&+&\frac{1}{2}\,(8\pi a)^2\int\frac{d^d{\bf k}}{(2\pi)^d}\,\int\frac{d^d{\bf q}}{(2\pi)^d}\,
\frac{1}{({\bf k}^2+\varepsilon^2)^2
  ({\bf k}^2+{\bf k}{\bf q}+{\bf q}^2+\varepsilon^2)}
+\cdots\, ,
\en
where the ellipses stand for more bubbles (all of them are regular at threshold and are
thus omitted here) and $d$ denotes the dimension of space.
Carrying out calculations, up to the terms that vanish at $\varepsilon\to 0$, one obtains:
\eq
S_1^d=\frac{2a}{\varepsilon}
-\frac{\sqrt{3}(\mu^2)^{d-3}a^2}{\pi}\,
\biggl[\frac{1}{3-d}+\Gamma'(1)+\ln 4\pi-\ln\frac{\varepsilon^2}{\mu^2}+1-\ln 2+\frac{1}{2}\,\ln 3-\frac{\sqrt{3}\pi}{9}\biggr]\, .
\en
On the other hand, carrying out calculations in the cutoff regularization, one obtains:
\eq\label{eq:S1_L}
S_1^\Lambda=-8\pi a^2\int^\Lambda\frac{d^3{\bf k}}{(2\pi)^3}\,
\frac{1}{({\bf k}^2+\varepsilon^2)^2}\,\sqrt{\frac{3}{4}\,{\bf k}^2+\varepsilon^2}
=\frac{2a}{\varepsilon}
-\frac{\sqrt{3}a^2}{\pi}\,\biggl[-\ln\frac{\varepsilon^2}{\Lambda^2}
+\ln 3-\frac{\sqrt{3}\pi}{9}-1\biggr]\, .
\en
Subtracting the above two equations, we get
\eq
S_1^d-S_1^\Lambda\simeq -\frac{\sqrt{3}(\mu^2)^{d-3}a^2}{\pi}\,
\biggl[\frac{1}{3-d}-\ln\frac{\Lambda^2}{\mu^2}+2.711355256\biggr]\, .
\en
Here, one encounters a subtlety: in Ref.~\cite{Beane}, a different
cutoff regularization is used at an intermediate step.
Namely, instead of Eq.~(\ref{eq:S1_L}), one writes:
\eq
\tilde S_1^\Lambda&=&-8\pi a\int^\Lambda\frac{d^3{\bf k}}{(2\pi)^3}\,\frac{1}{{\bf k}^2+\varepsilon^2}
\nonumber\\[2mm]
&+&\frac{1}{2}\,(8\pi a)^2\int^\Lambda\frac{d^3{\bf k}}{(2\pi)^3}\,
\int^\Lambda\frac{d^3{\bf q}}{(2\pi)^3}\,
\frac{1}{({\bf k}^2+\varepsilon^2)^2
  ({\bf k}^2+{\bf k}{\bf q}+{\bf q}^2+\varepsilon^2)}
+\cdots\, .
\en
Carrying out explicit calculations, it can be shown that\footnote{This result is interesting by its own. The above two expressions differ only in the ultraviolet region. In the quantity $S_1^\Lambda$, the limit $\Lambda\to\infty$ has been already
performed in the internal bubble (as it is implied by the form of integral equations we are
using), whereas in the quantity $\tilde S_1^\Lambda$ both cutoffs are held finite and equal.
As expected, the choice of different prescriptions
can be compensated by a finite renormalization of the coupling constant $H_0(\Lambda)$.}
\eq
\tilde S_1^\Lambda-S_1^\Lambda
=-\frac{\sqrt{3}a^2}{\pi}\,\biggl[\frac{\sqrt{3}}{2\pi}\,\ln{3}+\frac{1}{2}
-\frac{2}{\sqrt{3}\pi}+\frac{2}{\sqrt{3}\pi}\,\int_0^1 dk\ln k\biggl(\frac{k+2}{k^2+k+1}
-\frac{k-2}{k^2-k+1}\biggr)\biggr]\, ,
\en
and, therefore,
\eq\label{eq:3.65}
S_1^d-\tilde S_1^\Lambda\simeq -\frac{\sqrt{3}(\mu^2)^{d-3}a^2}{\pi}\,
\biggl[\frac{1}{3-d}-2\ln N +2\ln(\mu L)-0.020508246\biggr]\, .
\en
Comparing with Ref.~\cite{Beane}, one has to use the same prescription
as there. Hence, Eq.~(\ref{eq:3.65}) contains
two correction factors: the one accounting for the
different UV cutoffs used, and another that takes into account the difference between
the cutoff and the dimensional regularizations.

Further, the analog of our quantity ${\cal R}$ in Ref.~\cite{Beane} is defined as
\eq
\frac {1}{L^6}\sum_{{\bf k}\neq{\bf 0}}^\Lambda\sum_{\bf q}^\Lambda
\frac{1}{{\bf k}^4({\bf k}^2+{\bf q}^2+({\bf k}+{\bf q})^2)}
-\frac {1}{L^3}\sum_{{\bf k}\neq{\bf 0}}^\Lambda\frac{1}{{\bf k}^4}\int^\Lambda\frac{d^3{\bf q}}{2{\bf q}^2}
=-\frac{\sqrt{3}}{32\pi^3}\,\ln N-\frac{1}{(2\pi)^6}\, {\cal R}_1\, ,
\en
where $N$ is the same as in Eq.~(\ref{eq:tildeL}). The above expression can be further rewritten as
\eq\label{eq:R1}
-\frac{\sqrt{3}}{32\pi^3}\,\ln N-\frac{1}{(2\pi)^6}\, {\cal R}_1
&=&\frac{1}{L^3}\sum_{{\bf k}\neq{\bf 0}}^\Lambda\frac{1}{{\bf k}^4}
\biggl(\frac{1}{L^3}\,\sum_{\bf q}^\Lambda-\int^\Lambda\frac{d^3{\bf q}}{(2\pi)^3}\biggr)
\frac{1}{2}\,\frac{1}{{\bf k}^2+{\bf k}{\bf q}+{\bf q}^2}
\nonumber\\[2mm]
&+&\frac{1}{2L^3}\sum_{{\bf k}\neq{\bf 0}}^\Lambda\frac{1}{{\bf k}^4}
\int^\Lambda\frac{d^3{\bf q}}{(2\pi)^3}\,
\biggl(\frac{1}{{\bf k}^2+{\bf k}{\bf q}+{\bf q}^2}-\frac{1}{{\bf q}^2}\biggr)\, .
\en
When $\Lambda\to\infty$, the first term converges to $\dfrac{1}{2(2\pi)^6}\,{\cal G}$,
where ${\cal G}$ is given by Eq.~(\ref{eq:tildeG}). The evaluation of the integral
in the second term gives:
\eq
\frac{1}{2}\,\int d^3{\bf q}
\biggl(\frac{1}{{\bf k}^2+{\bf k}{\bf q}+{\bf q}^2}-\frac{1}{{\bf q}^2}\biggr)
&=&-\sqrt{3}\pi^2|{\bf k}|+\biggl\{
\biggl(\frac{\pi\Lambda^2}{|{\bf k}|}+\frac{\pi|{\bf k}|}{2}\biggr)
\,\ln\frac{\Lambda^2+\Lambda|{\bf k}|+{\bf k}^2}
{\Lambda^2-\Lambda|{\bf k}|+{\bf k}^2}-2\pi\Lambda
\nonumber\\[2mm]
&+&\sqrt{3}\pi|{\bf k}|\biggl(\arctan\frac{\sqrt{3}|{\bf k}|}{2\Lambda+|{\bf k}|}
+\arctan\frac{\sqrt{3}|{\bf k}|}{2\Lambda-|{\bf k}|}\biggr)\biggr\}\, .
\en
One has to insert this expression into Eq.~(\ref{eq:R1}) and evaluate the sum
over ${\bf k}$. Substituting the first term yields $-\dfrac{\sqrt{3}}{128\pi^4}\,({\cal L}
+2\pi\ln N)$, where ${\cal L}$ is given by Eq.~(\ref{eq:tildeL}).
Substituting the rest and performing the limit $\Lambda\to\infty$, one sees that the sum
ever ${\bf k}$ can be replaced by integration in this case. So, finally we get:
\eq
-\frac{\sqrt{3}}{32\pi^3}\,\ln N-\frac{1}{(2\pi)^6}\, {\cal R}_1
=\frac{1}{2(2\pi)^6}({\cal G}-\sqrt{3}\pi^2{\cal L})-\frac{\sqrt{3}}{32\pi^3}\,
\ln N-\frac{1}{(2\pi)^6}\,\chi\, ,
\en
where
\eq
\chi=\pi^2\biggl\{\frac{3}{2}\,\ln(3)-2+\frac{\sqrt{3}\pi}{2}+2\int dx\ln(x)
\biggl(\frac{x+2}{1+x+x^2}-\frac{x-2}{1-x+x^2}\biggr)\biggr\}\simeq -50.691124\, .
\en
>From this we finally get ${\cal R}_1=\dfrac{1}{2}\,({\cal G}-\sqrt{3}\pi^2{\cal L})-\chi$ and
\eq
{\cal R}\mbox{ [Ref.~\cite{Beane}]}\simeq {\cal R}_1+\sqrt{3}\pi^3\times 0.020508246
\simeq 19.186903\, ,
\en
which agrees with the value given in Ref.~\cite{Beane}.

Next, we calculate the term with two iterations of the driving term $Z$, which is an
infinite-volume counterpart of the quantity $S_2$, considered in
section~\ref{sec:two-iterations}. In dimensional regularization, we have
\eq
S_2^d=(8\pi a)^2
\int\frac{d^d{\bf k}}{(2\pi)^d}\,\int\frac{d^d{\bf q}}{(2\pi)^d}\,
\frac{1}{({\bf k}^2+\varepsilon^2)
  ({\bf k}^2+{\bf k}{\bf q}+{\bf q}^2+\varepsilon^2)
({\bf k}^2+\varepsilon^2)}
+\cdots\, ,
\en
where the ellipses, as before, stand for the non-singular terms at $\varepsilon\to 0$.
Evaluating this expression, we obtain
\eq
S_2^d=\frac{a^2(\mu^2)^{d-3}}{2\pi}\,
\biggl\{\frac{8\pi}{3}\,\biggl(\frac{1}{3-d}+\Gamma'(1)+\ln 4\pi-\ln\frac{\varepsilon^2}{\mu^2}\biggr)+{\cal J}_d\biggr\}\, ,
\en
where
\eq
{\cal J}_d=\int_0^1 xdx\biggl\{\frac{2\ln(g)+4}{(\frac{3}{2}-x)\sqrt{g}}-\frac{8}{x(\frac{3}{2}-x)}\,\arctan\frac{x}{2\sqrt{g}}\biggr\} \simeq-9.436982652\, ,
\quad
g=\frac{3-2x-x^2}{4}\, .
\en
The counterpart of this integral in the cutoff regularization has already been calculated
in section~\ref{sec:two-iterations}. Using this expression, we have
\eq
S_2^d-S_2^\Lambda&=&\frac{4a^2(\mu^2)^{d-3}}{3}\,
\biggl\{\frac{1}{3-d}+\Gamma'(1)-\ln\frac{\Lambda^2}{\mu^2}
+\frac{3}{8\pi}\,{\cal J}_d-\frac{3}{4\pi^4}\,{\cal I}_0\biggr\}
\nonumber\\[2mm]
&\simeq&\frac{4a^2(\mu^2)^{d-3}}{3}\,\biggl\{\frac{1}{3-d}-2\ln N
+2\ln(\mu L)+0.044721273\biggr\}\, .
\en
This enables one to relate the constants
${\cal Q}$ that appear in this work and in Ref.~\cite{Beane}:
\eq
{\cal Q}\mbox{ [Ref.~\cite{Beane}]}\simeq{\cal Q}+\frac{2\pi^4}{3}\,0.044721273
\simeq-102.1556055\, .
\en
This value is quite close to the one quoted in Ref.~\cite{Beane}.

\section{Conclusions and Outlook}
\label{sec:concl}

\begin{itemize}

\item[i)]
  Using the quantization condition derived in Refs.~\cite{pang1,pang2,Doring:2018xxx},
  we derive the expressions for the energy shift of the three-particle ground-state
  energy level and the first excited level at order $L^{-6}$, and the particle-dimer
  ground state at order $L^{-3}$. This derivation provides a beautiful test of the
  formalism.
  
\item[ii)]
  We carefully compare the result for the ground state to the expressions given in
  Refs.~\cite{Hansen:2015zta,Hansen:2016fzj,Sharpe:2017jej} and Ref.~\cite{Beane}.
  It has been shown that, apart from the relativistic corrections, which are absent in
  the present paper and in Ref.~\cite{Beane}, all these expressions agree completely.
  The expressions for the excited levels and the particle-dimer level are new
  (the latter reproduces the well-known two-particle L\"uscher formula at order $L^{-3}$). 
  
\item[iii)]
  The calculations for the higher excited levels can be done similarly to the first excited
  level. Also, the projection onto the different irreps of the octahedral groups can
  be treated in the same fashion.
  
\item[iv)]
  A natural question arises here: how do our results change, if the higher-order two-body
  and three-body effective couplings, as well as mixing to the higher partial waves are
  taken into account? We argue that the final results, which are expressed through the
  threshold amplitude $\hat{\cal M}$, do not change at all. All the above effects are just
  included in this amplitude since, according to the power counting, none of them
  contributes up to and including order $L^{-6}$.
  
\item[v)]
  From our results it is seen that the three-body force (encoded, in particular,
  in the coupling $\dfrac{H_0(\Lambda)}{\Lambda^2}$) starts to contribute at NNNLO
  to the three-particle energy levels, but appears already at leading order in the
  particle-dimer energy levels. Therefore, in order to measure, e.g., the three-pion
  force, it might be convenient to look at the $\pi\sigma$ scattering at
  higher quark masses, where $\sigma$ is a stable particle. With the use of
  a chiral extrapolation, one may then estimate the value of the three-pion coupling constant
  in the physical world.

    \end{itemize}

\acknowledgements
The authors would like to thank M. Hansen, F. Romero-L\'{o}pez, S. Sharpe,
C. Thomas and C. Urbach for interesting discussions. We would also like to thank
S. Beane, W. Detmold and M. Savage for providing us with the notes of
their calculations which made a detailed comparison of the results possible.
HWH and AR are grateful to the Mainz Institute for Theoretical
Physics (MITP) for its hospitality
and its partial support during the completion of this work.
JYP, JW, UGM and AR acknowledge the support from the DFG
(CRC 110 ``Symmetries and the Emergence of Structure in
QCD''). UGM and AR acknowledge support from
Volkswagenstiftung under contract no. 93562. 
The work of AR was in part supported by Shota Rustaveli National Science Foundation
(SRNSF), grant no. DI-2016-26.
HWH was supported by the Deutsche Forschungsgemeinschaft (DFG,
German Research Foundation) - Projektnummer 279384907 - SFB 1245
and the Federal Ministry
of Education and Research (BMBF) under contract 05P18RDFN1.
The work of UGM is supported in part by The Chinese Academy of Sciences (CAS)
President's International Fellowship Initiative (PIFI) (grant no. 2018DM0034).

  \appendix

\section{Energy shift of three-particle scattering states}
\label{app:num-res}
The energy of the scattering states vanishes in the infinite
volume limit. We quote their finite volume energy $E$ in terms of the quantity
$\kappa^2=L^2mE/(2\pi)^2$.

\subsection{Ground state ($A_{1}^{+}$)}

The energy shift of the ground state (which resides in the $A_{1}^{+}$
irrep) is: 
\begin{align}
\kappa^{2}= & \frac{g_{0}}{L}\left(1+\frac{g_{1}}{L}+\frac{g_{2}}{L^{2}}+\frac{g_{3}}{L^{3}}\ln\frac{mL}{2\pi}+\frac{g_{4}}{L^{3}}+\cdots\right)\,,
\end{align}
with 
\begin{align}
g_{0}= & \frac{3}{\pi}a\,,\nonumber \\
g_{1}= & 2.837297480\;a\,,\nonumber \\
g_{2}= & 9.725330808\;a^{2}\,,\nonumber \\
g_{3}= & 8\pi\left(\frac{2\sqrt{3}}{\pi}-\frac{8}{3}\right)a^{3}\,,\nonumber \\
  g_{4}= & \left(-5.159159617
           +6\pi\left(\frac{r}{a}\right)-8\pi\left(\frac{\hat{\mathcal{M}}}{a^{2}}\right)\right)a^{3}\,.
\end{align}

\subsection{The 1st excited state ($A_{1}^{+}$)}

The energy shift of the 1st excited state in the $A_{1}^{+}$ irrep is

\begin{align}
\kappa^{2}-1= & \frac{h_{0}}{L}\left(1+\frac{h_{1}}{L}+\frac{h_{2}}{L^{2}}+\frac{h_{3}}{L^{3}}\ln\frac{mL}{2\pi}+\frac{h_{4}}{L^{3}}+\cdots\right)\,,
\end{align}
with 
\begin{align}
h_{0}= & \frac{10}{\pi}a\,,\nonumber \\
h_{1}= & 0.279070\;a\,,\nonumber \\
h_{2}= & \left(8.494802+\frac{7\pi^{2}}{5}\left(\frac{r}{a}\right)\right)a^{2}
\,,\nonumber \\
h_{3}= & \frac{27}{5}\times8\pi\left(\frac{2\sqrt{3}}{\pi}-\frac{8}{3}\right)a^{3}\,,\nonumber \\
h_{4}= & \left(-172.001650+83.745841\left(\frac{r}{a}\right)-\frac{27}{5}\times8\pi\left(\frac{\hat{\mathcal{M}}}{a^{2}}\right)\right)a^{3}\,.
\end{align}

\subsection{The 1st excited state ($E^{+}$)}

The energy shift of the 1st excited state in the $E^{+}$ irrep is

\begin{align}
\kappa^{2}-1= & \frac{h_{0}^{'}}{L}\left(1+\frac{h_{1}^{'}}{L}+\frac{h_{2}^{'}}{L^{2}}+\frac{h_{4}^{'}}{L^{3}}+\cdots\right)\,,
\end{align}
with 
\begin{align}
h_{0}^{'}= & \frac{4}{\pi}a\,,\nonumber \\
h_{1}^{'}= & 2.984094\;a\,,\nonumber \\
h_{2}^{'}= & \left(3.001706+\frac{\pi^{2}}{2}\left(\frac{r}{a}\right)\right)a^{2}\,,\nonumber \\
h_{4}^{'}= & \left(-28.89478538+54.584571\left(\frac{r}{a}\right)\right)a^{3}\,.
\end{align}

\section{Transformations of sums and integrals}
\label{app:integrals}
\subsection{Derivation of Eq.~(\ref{eq:tildeL})}
The quantity ${\cal L}$ from Eq.~(\ref{eq:tildeL}) can be rewritten as
\eq
{\cal L}&=&\sum_{{\bf j}\neq{\bf 0}}\biggl(\frac{1}{{\bf j}^3}-\frac{1}{({\bf j}^2+1)^{3/2}}-\frac{3}{2}\,\frac{1}{({\bf j}^2+1)^{5/2}}-\frac{15}{8}\,\frac{1}{({\bf j}^2+1)^{7/2}}\biggr)
\nonumber\\[2mm]
&+&\lim_{N\to\infty}\biggl(\sum_{\bf j}^N\frac{1}{({\bf j}^2+1)^{3/2}}-4\pi\ln N-1\biggr)
+\frac{3}{2}\,\biggl(\sum_{\bf j}\frac{1}{({\bf j}^2+1)^{5/2}}-1\biggr)
\nonumber\\[2mm]
&+&\frac{15}{8}\,\biggl(\sum_{\bf j}\frac{1}{({\bf j}^2+1)^{7/2}}-1\biggr)\, ,
\en
where we have already performed the limit $N\to\infty$ in the first and the third terms. Using the Poisson summation formula in the second and the third terms, we
finally obtain
 \eq
{\cal L}&=&\sum_{{\bf j}\neq{\bf 0}}\biggl(\frac{1}{{\bf j}^3}-\frac{1}{({\bf j}^2+1)^{3/2}}-\frac{3}{2}\,\frac{1}{({\bf j}^2+1)^{5/2}}-\frac{15}{8}\,\frac{1}{({\bf j}^2+1)^{7/2}}\biggr)
+\pi(4\ln 2-1)-\frac{35}{8}
\nonumber\\[2mm]
&+&4\pi\sum_{{\bf j}\neq{\bf 0}}\biggl(K_0(2\pi|{\bf j}|)
+\pi|{\bf j}|K_1(2\pi|{\bf j}|)\biggr)
+2\pi^2\sum_{{\bf j}\neq{\bf 0}}\biggl(K_1(2\pi|{\bf j}|)
+\pi|{\bf j}|K_0(2\pi|{\bf j}|)\biggr)\, ,
\en
where $K_0(z),K_1(z)$ denote the modified Bessel functions of the second kind.
Carrying out the summation in the above equation numerically, we finally arrive
at the value given in Eq.~(\ref{eq:tildeL}).

\subsection{Derivation of Eq.~(\ref{eq:tildeG})}
Using the Poisson summation formula, we get
\eq
{\cal I}({\bf n})=\pi\sum_{{\bf j}\neq{\bf 0}}\frac{e^{i\pi{\bf n}{\bf j}}}{|{\bf j}|}\,\exp\bigl(-\sqrt{3}\pi|{\bf j}|\,|{\bf n}| \bigr)\, .
\en
Carrying out the summation numerically, we arrive at the value given in
Eq.~(\ref{eq:tildeG}). 

\subsection{Derivation of Eq.~(\ref{eq:S21})}

  Consider the quantity
  \eq
  \Sigma=\frac{1}{L^6}\,\sum_{{\bf k}\neq{\bf 0}}^\Lambda
  \sum_{{\bf q}\neq{\bf 0}}^\Lambda \frac{1}{{\bf k}^2}\,
  \frac{1}{{\bf k}^2+{\bf q}^2+{\bf k}{\bf q}}\,
  \frac{1}{{\bf q}^2}
  =\frac{1}{(2\pi)^6}\,\sum_{{\bf j}\neq{\bf 0}}^N
  \sum_{{\bf n}\neq{\bf 0}}^N \frac{1}{{\bf j}^2}\,
  \frac{1}{{\bf j}^2+{\bf n}^2+{\bf j}{\bf n}}\,
  \frac{1}{{\bf n}^2}\, ,
  \en
  where ${\bf j},{\bf n}\in\mathbb{Z}^3$ and $N=\dfrac{\Lambda L}{2\pi}$.
  We may further write $\Sigma=(\Sigma-\hat\Sigma)+\hat\Sigma$, where
  \eq
  \hat\Sigma=\frac{1}{(2\pi)^6}\int_1^N\frac{d^3{\bf j}}{{\bf j}^2}
  \int_1^N\frac{d^3{\bf n}}{{\bf n}^2}\, \frac{1}{{\bf j}^2+{\bf n}^2+{\bf j}{\bf n}}\, .
  \en
  After angular integration, this integral takes the form
  \eq
  \hat\Sigma=\frac{1}{4\pi^4}\,\int_1^N\int_1^N\frac{dxdy}{xy}\,\ln\frac{x^2+y^2+xy}{x^2+y^2-xy}\, .
  \en
  Using partial integration, it is possible to perform the integral over the variable $y$ analytically.
  The result is given by:
  \eq
  \hat\Sigma=\frac{\ln N}{2\pi^4}\,\int_{1/N}^1\frac{dx}{x}\,\ln\frac{x^2+x+1}{x^2-x+1}
  +\frac{1}{2\pi^4}\,\int_{1/N}^1\frac{dx}{x}\,\ln x\ln\frac{x^2+x+1}{x^2-x+1}\, .
  \en
  For large values of $N$, one may replace $1/N$ by zero. The first integral then
  equals $\pi^2/6$, and we get
  \eq
  \hat\Sigma=\frac{1}{12\pi^2}\,\ln N+\frac{1}{2\pi^4}\,
  \int_0^1\frac{dx}{x}\,\ln x\ln\frac{x^2+x+1}{x^2-x+1}
  =\frac{1}{12\pi^2}\,\ln N+\mbox{\rm const}\, .
  \en
  The constant does not play any role and can be dropped. Performing the limit
  $N\to\infty$ in the quantity $\Sigma-\hat\Sigma$, we finally arrive at Eq.~(\ref{eq:S21}).

\subsection{Derivation of Eq.~(\ref{eq:calQ})}

The infinite sum that enters Eq.~(\ref{eq:calQ}) can be rewritten in the following way:
\eq\label{eq:didi0}
\Sigma&=&\sum_{{\bf j}\neq{\bf 0}}^N\sum_{{\bf n}\neq{\bf 0}}^N
\frac{1}{{\bf j}^2}\,\frac{1}{{\bf j}^2+{\bf j}{\bf n}+{\bf n}^2}\,
\frac{1}{{\bf n}^2}
\nonumber\\[2mm]
&=&
\sum_{{\bf j}\neq{\bf 0}}^N\sum_{{\bf n}\neq{\bf 0}}^N
\frac{1}{{\bf j}^2({\bf j}^2+1)^3}\,\frac{1}{{\bf j}^2+{\bf j}{\bf n}+{\bf n}^2}\,
\frac{1}{{\bf n}^2({\bf n}^2+1)^3}
\nonumber\\[2mm]
&+&2\sum_{{\bf j}\neq{\bf 0}}^N\sum_{{\bf n}\neq{\bf 0}}^N\varphi({\bf j})
\frac{1}{({\bf j}^2+{\bf j}{\bf n}+{\bf n}^2)({\bf j}^2+{\bf j}{\bf n}+{\bf n}^2+1)^3}\,
\frac{1}{{\bf n}^2({\bf n}^2+1)^3}
\nonumber\\[2mm]
&+&2\sum_{{\bf j}\neq{\bf 0}}^N\sum_{{\bf n}\neq{\bf 0}}^N\varphi({\bf j})h({\bf j},{\bf n})\frac{1}{{\bf n}^2({\bf n}^2+1)^3}
\nonumber\\[2mm]
&+&\sum_{{\bf j}\neq{\bf 0}}^N\sum_{{\bf n}\neq{\bf 0}}^N\varphi({\bf j})\frac{1}{({\bf j}^2+{\bf j}{\bf n}+{\bf n}^2)({\bf j}^2+{\bf j}{\bf n}+{\bf n}^2+1)^3}\,\varphi({\bf n})
\nonumber\\[2mm]
&+&\sum_{{\bf j}\neq{\bf 0}}^N\sum_{{\bf n}\neq{\bf 0}}^N\varphi({\bf j})h({\bf j},{\bf n})\varphi({\bf n})\, ,
\en
where
\eq
\varphi({\bf n})&=&\frac{1}{{\bf n}^2+1}+\frac{1}{({\bf n}^2+1)^2}+\frac{1}{({\bf n}^2+1)^3}\, ,
\nonumber\\[2mm]
h({\bf j},{\bf n})&=&\frac{1}{{\bf j}^2+{\bf j}{\bf n}+{\bf n}^2+1}
+\frac{1}{({\bf j}^2+{\bf j}{\bf n}+{\bf n}^2+1)^2}
+\frac{1}{({\bf j}^2+{\bf j}{\bf n}+{\bf n}^2+1)^3}\, .
\en
The first, second and fourth terms in Eq.~(\ref{eq:didi0}) can be grouped 
together. We denote the sum of these three terms by $\Sigma_1$.
Further, in the third and the fifth terms of the same equation, 
we add and subtract the terms with ${\bf j}={\bf 0}$ or ${\bf n}={\bf 0}$.
As a result, we obtain $\Sigma=\Sigma_1+\Sigma_2+\Sigma_3+\Sigma_4$,
where the individual terms take the following form
\eq\label{eq:didi1}
\Sigma_1&=&
\sum_{{\bf j}\neq{\bf 0}}^N\sum_{{\bf n}\neq{\bf 0}}^N\biggl\{
\frac{1}{{\bf j}^2({\bf j}^2+1)^3}\,\frac{1}{{\bf j}^2+{\bf j}{\bf n}+{\bf n}^2}\,
\frac{1}{{\bf n}^2({\bf n}^2+1)^3}
\nonumber\\[2mm]
&+&\varphi({\bf j})\frac{1}{({\bf j}^2+{\bf j}{\bf n}+{\bf n}^2)({\bf j}^2+{\bf j}{\bf n}+{\bf n}^2+1)^3}\,\frac{1}{{\bf n}^2}\biggl(1+\frac{1}{({\bf n}^2+1)^3}\biggr)\biggr\}\, ,
\nonumber\\[2mm]
\Sigma_2&=&-6\sum_{{\bf n}\neq{\bf 0}}^N\frac{1}{{\bf n}^2}\,\varphi({\bf n})-27\, ,
\nonumber\\[2mm]
\Sigma_3&=&2\sum_{{\bf n}\neq{\bf 0}}^N\sum_{\bf j}^N
\varphi({\bf j})h({\bf j},{\bf n})\frac{1}{{\bf n}^2({\bf n}^2+1)^3}\, ,
\nonumber\\[2mm]
\Sigma_4&=&\sum_{\bf n}^N\sum_{\bf j}^N\varphi({\bf j})h({\bf j},{\bf n})
\varphi({\bf n})\, .
\en
$\Sigma_1$ and $\Sigma_2$ are rapidly converging sums. We may perform the limit
$N\to\infty$ in these terms, obtaining
\eq
\Sigma_1\simeq 2.8391985\, ,
\quad\quad \Sigma_2\simeq -120.8804992739\, .
\en
In order to calculate $\Sigma_3$ and $\Sigma_4$, the Poisson summation method
is used. The quantity $\Sigma_3$ is equal to
\eq
\Sigma_3= 2\sum_{\bf n}\frac{1}{{\bf n}^2({\bf n}^2+1)^3}\,\Psi({\bf n})\, ,
\en
where
\eq
\Psi({\bf n})=\sum_{\bf j}\int d^3{\bf k}e^{2\pi i{\bf k}{\bf j}}\varphi({\bf k})
h({\bf k},{\bf n})=\sum_{a=1}^3\sum_{b=1}^3I_{ab}\, ,
\en
and
\eq
I_{ab}&=&\int d^3{\bf k}e^{2\pi i{\bf k}{\bf j}}\frac{1}{({\bf k}^2+1)^a
({\bf k}^2+{\bf k}{\bf n}+{\bf n}^2+1)^b}
\nonumber\\[2mm]
&=&\frac{\Gamma(a+b)}{\Gamma(a)\Gamma(b)}
\int_0^1dxe^{-i\pi x{\bf n}{\bf j}}\int d^3{\bf k}e^{2\pi i{\bf k}{\bf j}}
\frac{x^{b-1}(1-x)^{a-1}}{(f^2+{\bf k}^2)^{a+b}}
\nonumber\\[2mm]
&=&\frac{1}{\Gamma(a)\Gamma(b)}
\int_0^1dxe^{-i\pi x{\bf n}{\bf j}}x^{b-1}(1-x)^{a-1}\int_0^\infty d\lambda
\lambda^{a+b-1}\int d^3{\bf k}e^{2\pi i{\bf k}{\bf j}-\lambda(f^2+{\bf k}^2)}
\nonumber\\[2mm]
&=&\frac{\pi^{3/2}}{\Gamma(a)\Gamma(b)}
\int_0^1dx\cos(\pi x{\bf n}{\bf j})x^{b-1}(1-x)^{a-1}\int_0^\infty d\lambda
\lambda^{a+b-5/2}\int e^{-\lambda f^2-\pi^2{\bf j}^2/\lambda}\, .
\en
Here,
\eq
f^2=1+\biggl(x-\frac{x^2}{4}\biggr){\bf n}^2\, .
\en
Note that we have set $N\to\infty$ everywhere, since all integrals and sums
are convergent. Using the above equations, we finally get:
\eq
\Sigma_3&=&2\pi^{3/2}\sum_{{\bf n}\neq{\bf 0}}\sum_{\bf j}\frac{1}{{\bf n}^2({\bf n}^2+1)^3}
\int_0^1dx\cos(\pi x{\bf n}{\bf j})\int_0^\infty \frac{d\lambda}{\sqrt{\lambda}}
  e^{-\lambda f^2-\pi^2{\bf j}^2/\lambda}
\nonumber\\[2mm]
&\times&\biggl(1+\lambda x+\frac{\lambda^2x^2}{2}\biggr)
\biggl(1+\lambda (1-x)+\frac{\lambda^2(1-x)^2}{2}\biggr)\, .
\en
Evaluating this expression numerically, we get
\eq
\Sigma_3\simeq 27.2953+0.9312\simeq 28.2266\, ,
\en
where the first and the second terms correspond to ${\bf j}={\bf 0}$ and 
${\bf j}\neq{\bf 0}$ in the sum over all ${\bf j}$.

We use the same technique to calculate $\Sigma_4$. The result is given by:
\eq
\Sigma_4=\sum_{\bf n}\sum_{\bf j}\sum_{a=1}^3\sum_{b=1}^3\sum_{c=1}^3 I_{abc}\, ,
\en
where
\eq
I_{abc}&=&\int^N\int^N d^3{\bf q}d^3{\bf k}
e^{2\pi i({\bf j}{\bf k}+{\bf q}{\bf n})}
\frac{1}{({\bf q}^2+1)^a({\bf k}^2+1)^b({\bf q}^2+{\bf q}{\bf k}+{\bf k}^2+1)^c}
\nonumber\\[2mm]
&=&\frac{\Gamma(a+b+c)}{\Gamma(a)\Gamma(b)\Gamma(c)}\,
\int_0^\infty dxdydz\delta(1-x-y-z) x^{a-1}y^{b-1}(1-x-y)^{c-1}
\nonumber\\[2mm]
&\times&\int^N\int^N d^3{\bf q}d^3{\bf k}\frac{e^{2\pi i({\bf j}{\bf k}+{\bf q}{\bf n})}}
{(1+(1-y){\bf q}^2+(1-x){\bf k}^2+(1-x-y){\bf q}{\bf k})^{a+b+c}}\, .
\en
The above expression is finite in the limit $N\to\infty$, except the term with
$a=b=c=1$ and ${\bf n}={\bf j}=0$. We can single out this divergent term and
calculate the rest, where the limits of integration can be set to infinity.
Performing the change of the variables
\eq
{\bf q}=\frac{1}{\sqrt{1-y}}\,\biggl(\frac{{\bf u}}{\sqrt{2(1+\alpha)}}
+\frac{{\bf v}}{\sqrt{2(1-\alpha)}}\biggr)\, ,\quad\quad
{\bf k}=\frac{1}{\sqrt{1-x}}\,\biggl(\frac{{\bf u}}{\sqrt{2(1+\alpha)}}
-\frac{{\bf v}}{\sqrt{2(1-\alpha)}}\biggr)\, ,
\en
where
\eq
\alpha=\frac{1-x-y}{2\sqrt{1-x}\sqrt{1-y}}\, ,
\en
we get
\eq
I_{abc}&=&\frac{\Gamma(a+b+c)}{\Gamma(a)\Gamma(b)\Gamma(c)}\,
\int_0^\infty dxdydz\delta(1-x-y-z) \frac{x^{a-1}y^{b-1}(1-x-y)^{c-1}}
{((1-x)(1-y)(1-\alpha^2))^{3/2}}
\nonumber\\[2mm]
&\times&\int d^6{\bf Q}\frac{e^{2\pi i{\bf Q}{\bf R}}}{(1+{\bf Q}^2)^{a+b+c}}\, ,
\en
where we have defined the vectors in the 6-dimensional space
\eq
{\bf Q}&=&({\bf u},{\bf v})\, ,\quad\quad
\nonumber\\[2mm]
{\bf R}&=&\biggl(
\frac{1}{\sqrt{2(1+\alpha)}}\biggl(\frac{{\bf j}}{\sqrt{1-x}}
+\frac{\bf n}{\sqrt{1-y}}\biggr),
\frac{1}{\sqrt{2(1-\alpha)}}\biggl(\frac{{\bf j}}{\sqrt{1-x}}
-\frac{\bf n}{\sqrt{1-y}}\biggr)\biggr)\, .
\en
Carrying out the integration over $d^6{\bf Q}$ by exponentiating the denominator, we get
\eq
I_{abc}&=&\frac{\pi^3}{\Gamma(a)\Gamma(b)\Gamma(c)}
\int_0^\infty dxdydz\delta(1-x-y-z) \frac{x^{a-1}y^{b-1}(1-x-y)^{c-1}}
{((1-x)(1-y)(1-\alpha^2))^{3/2}}
\nonumber\\[2mm]
&\times&\int d\lambda\lambda^{a+b+c-4}\exp(-\lambda-\pi^2{\bf R}^2/\lambda)\, ,
\en
where
\eq
{\bf R}^2=\biggl(\frac{{\bf j}^2}{1-x}+\frac{{\bf n}^2}{1-y}\biggr)\frac{1}{1-\alpha^2}
-\frac{2\alpha{\bf j}{\bf n}}{\sqrt{1-x}\sqrt{1-y}(1-\alpha^2)}\, .
\en
The infinite-volume integral is considered in Eq.~(\ref{eq:tI0}).
Adding everything together, we get
\eq
\lim_{N\to\infty}\biggl(\Sigma_4-\frac{8\pi^4}{3}\,\ln N\biggr)&=&
\sum_{\bf j}\sum_{\bf n}(1-\delta_{{\bf j}{\bf 0}}\delta_{{\bf n}{\bf 0}})
\int_0^\infty dxdydz\delta(1-x-y-z) 
\nonumber\\[2mm]
&\times&\int_0^\infty\frac{d\lambda}{\lambda}\,
\frac{\pi^3\exp(-\lambda-\pi^2{\bf R}^2/\lambda)}
{((1-x)(1-y)(1-\alpha^2))^{3/2}}
\biggl(1+\lambda x+\frac{\lambda^2x^2}{2}\biggr)
\nonumber\\[2mm]
&\times&\biggl(1+\lambda y+\frac{\lambda^2y^2}{2}\biggr)
\biggl(1+\lambda (1-x-y)+\frac{\lambda^2(1-x-y)^2}{2}\biggr)
\nonumber\\[2mm]
&+&\int_0^\infty dxdydz\delta(1-x-y-z) 
\int_0^\infty\frac{d\lambda}{\lambda}\,
\frac{\pi^3\exp(-\lambda)}
{((1-x)(1-y)(1-\alpha^2))^{3/2}}
\nonumber\\[2mm]
&\times&\biggl\{\biggl(1+\lambda x+\frac{\lambda^2x^2}{2}\biggr)
\biggl(1+\lambda y+\frac{\lambda^2y^2}{2}\biggr)
\nonumber\\[2mm]
&\times&\biggl(1+\lambda (1-x-y)+\frac{\lambda^2(1-x-y)^2}{2}\biggr)-1\biggr\}
+{\cal I}_0
\nonumber\\[2mm]
&\simeq& -150.15376281+25.1433615+4.705512\simeq -120.304889\, .
\en
Here, the first, second and the third terms correspond to:
${\bf j}={\bf n}={\bf 0}$; either ${\bf j}={\bf 0}$ or  ${\bf n}={\bf 0}$;
both ${\bf j},{\bf n}\neq {\bf 0}$.

Adding everything, we arrive at the numerical value given in Eq.~(\ref{eq:calQ}).

\subsection{Derivation of Eq.~(\ref{eq:tI0})}

Carrying out the integration over the angles and rescaling the integration
variables, the l.h.s. of Eq.~(\ref{eq:tI0}) takes the form
\eq
I_\varepsilon=\frac{a^2}{8\pi^4}\,\int_0^x\frac{pdp}{p^2+1}\,
\int_0^x\frac{qdq}{q^2+1}\,\ln\frac{p^2+pq+q^2+1}{p^2-pq+q^2+1}\, ,
\en
where $x=\Lambda/\varepsilon$. Performing the partial integration and rescaling
the integration variables in some terms, we arrive at
\eq
I_\varepsilon&=& 
  \frac{a^2}{8\pi^4}\,\ln x\int_{x^{-1}}^1\frac{qdq}{q^2+x^{-2}}\,
\ln
\frac{q^2+q+1+x^{-2}}{q^2-q+1+x^{-2}}
\nonumber\\[2mm]
 &-&\frac{a^2}{16\pi^4}\,\int_0^xdp\ln(p^2+1)\int_0^x\frac{qdq}{q^2+1}\,
\biggl[\frac{2p+q}{p^2+pq+q^2+1}-\frac{2p-q}{p^2-pq+q^2+1}\biggr]\, .
\en
The integral over the variable $q$ in the second term can be performed
analytically. Neglecting everywhere the terms that vanish at $x\to\infty$, we
finally get:
\eq
I_\varepsilon=\frac{a^2}{24\pi^2}\,\ln x
+\frac{a^2}{8\pi^4}\,\int_0^1dp\frac{\ln p}{p}\,\ln\frac{p^2+p+1}{p^2-p+1}
-\frac{a^2}{16\pi^3}\,\int_0^\infty dp\frac{\ln(p^2+1)}{p^2+1}\,\biggl[
1-\frac{1}{\sqrt{4+3p^2}}\biggr]\, .
\en
>From this expression, one can read off the expression for the numerical
constant ${\cal I}_0$ in Eq.~(\ref{eq:tI0}):
\eq
{\cal I}_0=8\pi^2\int_0^1dp\frac{\ln p}{p}\,\ln\frac{p^2+p+1}{p^2-p+1}
-4\pi^3\int_0^\infty dp\frac{\ln(p^2+1)}{p^2+1}\,\biggl[
1-\frac{1}{\sqrt{4+3p^2}}\biggr]\simeq -375.754658\, .\hspace*{.3cm} 
\en

\section{Infinite sums}
\label{app:sums}

In this appendix, we collect numerical values for all infinite sums over integer numbers,
which appear in the main text. In the following, ${\bf j}_0=(0,0,1)$ is a fixed unit vector,
and the sum over the unit vector ${\bf i}_0$ runs over six possible orientations
on the cubic lattice. The method of calculation of some of these sums is considered in
Appendix~\ref{app:integrals}, and the remaining sums are calculated along the
same pattern.

To calculate the energy shift of the ground state, we need the sums over the single
integer vector:
\eq
{\cal I}&=&\lim_{N\to\infty}\biggr\{
\sum_{{\bf j}\neq {\bf 0}}^N\frac{1}{{\bf j}^2}-4\pi N\biggr\}\simeq-8.91363291781\, ,
\nonumber\\[2mm]
{\cal J}&=&\sum_{{\bf j}\neq {\bf 0}}^\infty\frac{1}{{\bf j}^4}\simeq 16.532315959\, ,
\nonumber\\[2mm]
{\cal K}&=&\sum_{{\bf j}\neq {\bf 0}}^\infty\frac{1}{{\bf j}^6}\simeq 8.401923974433\, ,
\en
and
\eq\label{eq:tildeL}
{\cal L}
=\lim_{N\to\infty}\biggl[\sum_{{\bf j}\neq{\bf 0}}^N\frac{1}{|{\bf j}|^3}
-4\pi\ln N\biggr]\simeq 3.82192350\, ,
\en
and
\eq\label{eq:tildeG}
{\cal G}=\sum_{{\bf n}\neq{\bf 0}}\frac{1}{{\bf n}^4}\,
  \tilde {\cal I}({\bf n})\simeq 0.123259025\, ,
\en
as well as the sums over two integer vectors:
\eq\label{eq:calQ}
{\cal Q}=\lim_{N\to\infty}\biggl[\frac{1}{2}\,\sum_{{\bf j}\neq{\bf 0}}^N\sum_{{\bf n}\neq{\bf 0}}^N\frac{1}{{\bf j}^2}\,\frac{1}{{\bf j}^2+{\bf j}{\bf n}+{\bf n}^2}\,\frac{1}{{\bf n}^2}
    -\frac{4\pi^4}{3}\,\ln N\biggr]\simeq -105.0597779\, .
\en
In the expansion of the energy shift of the first excited state in the $A_1^+$ irrep, the
following sums over a single integer vector appear (here, ${\bf j}_0$ denotes an integer
vector with $|{\bf j}_0|=1$):
\eq
{\cal I}_1&=&\lim_{N\to\infty}\biggl(\sum_{|{\bf j}|\neq 1}^N\frac{1}{{\bf j}^2-1}-4\pi N\biggr)\simeq -1.2113357\, ,
\nonumber\\[2mm]
{\cal J}_1&=&\sum_{|{\bf j}|\neq 1}\frac{1}{({\bf j}^2-1)^2}\simeq 23.2432219\, ,
\nonumber\\[2mm]
{\cal K}_1&=&\sum_{|{\bf j}|\neq 1}\frac{1}{({\bf j}^2-1)^3}\simeq 13.0593768\, ,
\nonumber\\[2mm]
{\cal I}_2&=&\lim_{N\to\infty}\biggl(\sum_{{\bf j}\neq{\bf 0},-{\bf j}_0}^N\frac{1}{{\bf j}^2+{\bf j}{\bf j}_0}-4\pi N\biggr)\simeq -6.37480912\, ,
\nonumber\\[2mm]
{\cal J}_2&=&\sum_{{\bf j}\neq{\bf 0},-{\bf j}_0}\frac{1}{({\bf j}^2+{\bf j}{\bf j}_0)^2}\simeq 18.343774\, ,
\nonumber\\[2mm]
{\cal K}_2&=&\sum_{{\bf j}\neq{\bf 0},-{\bf j}_0}\frac{1}{({\bf j}^2+{\bf j}{\bf j}_0)^3}\simeq 10.2376434\, ,
\en
and
\eq
{\cal J}_{11}&=&
\sum_{|{\bf j}|\neq 0,1}^\infty
\frac{1}{({\bf j}^2-1)^2}\simeq 22.2432219\, ,
\nonumber\\[2mm]
{\cal J}_{12}&=&
\sqrt{6}\sum_{|{\bf j}|\neq 0,1}^\infty
\frac{1}{{\bf j}^2-1}\, \frac{1}{{\bf j}^2+{\bf j}{\bf j}_0}\simeq 40.3648902\, ,
\nonumber\\[2mm]
{\cal J}_{22}&=&\sum_{|{\bf i}_0|=1}\sum_{|{\bf j}|\neq 0,1}^\infty
\frac{1}{{\bf i}_0{\bf j}+{\bf j}^2}\frac{1}{{\bf j}^2+{\bf j}{\bf j}_0} \simeq 76.3433459\, ,
\\[2mm]
{\cal K}_{11}&=&
\sum_{|{\bf j}|\neq 0,1}^\infty
\frac{1}{({\bf j}^2-1)^3}\simeq 14.0593768\, ,
\nonumber\\[2mm]
{\cal K}_{12}&=&\frac{\sqrt{6}}{2}\,\sum_{|{\bf j}|\neq 0,1}^\infty
\biggl(\frac{1}{({\bf j}^2-1)^2}\,\frac{1}{{\bf j}^2+{\bf j}{\bf j}_0}
+\frac{1}{{\bf j}^2-1}\,
\frac{1}{({\bf j}^2+{\bf j}{\bf j}_0)^2}\biggr)\simeq 19.5592825\, ,
\nonumber\\[2mm]
{\cal K}_{22}&=&\frac{1}{2}\,\sum_{|{\bf i}_0|=1}\sum_{|{\bf j}|\neq 0,1}^\infty
\biggl(\frac{1}{({\bf i}_0{\bf j}+{\bf j}^2)^2}\,
\frac{1}{{\bf j}^2+{\bf j}{\bf j}_0}
+\frac{1}{{\bf i}_0{\bf j}+{\bf j}^2}\,
\frac{1}{({\bf j}^2+{\bf j}{\bf j}_0)^2}\biggr)\simeq 27.4618302\, ,
\\[2mm]
{\cal L}_{11}&=&\lim_{N\to\infty}\biggl(\sum_{|{\bf j}|\neq 0,1}^N
\frac{\sqrt{{\bf j}^2-\dfrac{4}{3}}}{({\bf j}^2-1)^2}-4\pi\ln N\biggr)\simeq
7.153930983\, ,
\nonumber\\[2mm]
{\cal L}_{12}&=&\sqrt{6}\lim_{N\to\infty}\biggl(\sum_{|{\bf j}|\neq 0,1}^N
\frac{\sqrt{{\bf j}^2-\dfrac{4}{3}}}{({\bf j}^2-1)({\bf j}^2+{\bf j}{\bf j}_0)}-4\pi\ln N\biggr)\simeq 3.048537581\, ,
\nonumber\\[2mm]
{\cal L}_{22}&=&\sum_{|{\bf i}_0|=1}\lim_{N\to\infty}\biggl(\sum_{|{\bf j}|\neq 0,1}^N
\frac{\sqrt{{\bf j}^2-\dfrac{4}{3}}}{({\bf i}_0{\bf j}+{\bf j}^2)({\bf j}^2+{\bf j}{\bf j}_0)}-4\pi\ln N\biggr)\simeq -17.27264446\, ,
\\[2mm]
{\cal G}_{11}&=&\sum_{|{\bf j}|\neq 0,1}^\infty
\frac{{\cal I}(\bf j)}
{({\bf j}^2-1)^2}\simeq -1.028613861\, ,
\nonumber\\[2mm]
{\cal G}_{12}&=&\sqrt{6}\sum_{|{\bf j}|\neq 0,1}^\infty
\frac{{\cal I}(\bf j)}
{({\bf j}^2-1)({\bf j}^2+{\bf j}{\bf j}_0)}\simeq -1.5491996719\, ,
\nonumber\\[2mm]
{\cal G}_{22}&=&\sum_{|{\bf i}_0|=1}\sum_{|{\bf j}|\neq 0,1}^\infty
\frac{{\cal I}(\bf j)}
{({\bf i}_0{\bf j}+{\bf j}^2)({\bf j}^2+{\bf j}{\bf j}_0)}\simeq -2.3361490578\, ,
\en
as well as the sums over two integer vectors:
\eq
{\cal Q}_{11}&=&\lim_{N\to\infty}\biggl(\frac{1}{2}\,
\sum_{|{\bf j}|\neq 0,1}^N\sum_{|{\bf n}|\neq 0,1}^N
\frac{1}{{\bf j}^2-1}\,
\frac{1}{{\bf j}^2+{\bf j}{\bf n}+{\bf n}^2-1}\,
\frac{1}{{\bf n}^2-1}-\frac{4\pi^4}{3}\,\ln N\biggr)\simeq -39.5306\, ,
\nonumber\\[2mm]
{\cal Q}_{12}&=&\sqrt{6}\lim_{N\to\infty}\biggl(\frac{1}{2}\,
\sum_{|{\bf j}|\neq 0,1}^N\sum_{|{\bf n}|\neq 0,1}^N
\frac{1}{{\bf j}^2-1}\,
\frac{1}{{\bf j}^2+{\bf j}{\bf n}+{\bf n}^2-1}\,
\frac{1}{{\bf n}^2+{\bf n}{\bf j}_0}-\frac{4\pi^4}{3}\,\ln N\biggr)\simeq  -219.3579\, ,
\nonumber\\[2mm]
{\cal Q}_{22}&=&\sum_{|{\bf i}_0|=1}\lim_{N\to\infty}\biggl(\frac{1}{2}\,
\sum_{|{\bf j}|\neq 0,1}^N\sum_{|{\bf n}|\neq 0,1}^N
\frac{1}{{\bf i}_0{\bf j}+{\bf j}^2}\,
\frac{1}{{\bf j}^2+{\bf j}{\bf n}+{\bf n}^2-1}\,
\frac{1}{{\bf n}^2+{\bf n}{\bf j}_0}
-\frac{4\pi^4}{3}\,\ln N\biggr)
\simeq -783.596\, .
\nonumber\\
&&
\en

Finally, the sums over the single and two integer vectors, appearing in the
expression of the energy shift of the first excited level in the irrep $E^+$, are
equal to:
\eq
\label{eq:begin}
{\cal J}'_{22}&=&\sum_{|{\bf i}_0|=1}\lambda({\bf i}_0)\sum_{|{\bf j}|\neq 0,1}^\infty
\frac{1}{{\bf i}_0{\bf j}+{\bf j}^2}\frac{1}{{\bf j}^2+{\bf j}{\bf j}_0}\simeq 0.278429651\, ,
\\[2mm]
{\cal K}'_{22}&=&\frac{1}{2}\,\sum_{|{\bf i}_0|=1}\lambda({\bf i}_0)
\sum_{|{\bf j}|\neq 0,1}^\infty
\biggl(\frac{1}{({\bf i}_0{\bf j}+{\bf j}^2)^2}\,
\frac{1}{{\bf j}^2+{\bf j}{\bf j}_0}
+\frac{1}{{\bf i}_0{\bf j}+{\bf j}^2}\,
\frac{1}{({\bf j}^2+{\bf j}{\bf j}_0)^2}\biggr)\simeq 0.449738111\, ,
\\[2mm]
{\cal L}'_{22}&=&\sum_{|{\bf i}_0|=1}\lambda({\bf i}_0)
\sum_{|{\bf j}|\neq 0,1}^\infty
\frac{\sqrt{{\bf j}^2-\dfrac{4}{3}}}{({\bf i}_0{\bf j}+{\bf j}^2)({\bf j}^2+{\bf j}{\bf j}_0)}\simeq 0.2888875370\, ,
\\[2mm]
{\cal G}'_{22}&=&\sum_{|{\bf i}_0|=1}\lambda({\bf i}_0)\sum_{|{\bf j}|\neq 0,1}^\infty
\frac{{\cal I}_({\bf j})}{({\bf i}_0{\bf j}+{\bf j}^2)({\bf j}^2+{\bf j}{\bf j}_0)}\simeq -0.018431808\, ,
\\[2mm]
\label{eq:end}
{\cal Q}'_{22}&=&\frac{1}{2}\,\sum_{|{\bf i}_0|=1}\lambda({\bf i}_0)
\sum_{|{\bf j}|\neq 0,1}^\infty\sum_{|{\bf n}|\neq 0,1}^\infty
\frac{1}{{\bf i}_0{\bf j}+{\bf j}^2}\,
\frac{1}{{\bf j}^2+{\bf j}{\bf n}+{\bf n}^2-1}\,
\frac{1}{{\bf n}^2+{\bf n}{\bf j}_0}\simeq 0.07825\, .
\en
Here, 
\eq
\lambda({\bf i}_0)=
\left\{
\begin{array}{l c l}
1 &,& \mbox{\quad if}\quad{\bf i}_0=(0,0,\pm 1) \\
-\dfrac{1}{2} &,& \mbox{\quad otherwise}\\
\end{array} 
\right.\,,
\en
and the integer vector ${\bf j}$ again belongs to the shell $r$.
Note the absence of the logarithmic divergences
in Eqs.~(\ref{eq:begin})-(\ref{eq:end}).

\end{document}